\documentclass[11pt,letterpaper]{JHEP3}
\usepackage{array}
\usepackage{cite}
\input epsf.tex
\newcommand{\beq}{\begin{eqnarray}}
\newcommand{\eeq}{\end{eqnarray}}

\def\ltap{\ \raise.3ex\hbox{$<$\kern-.75em\lower1ex\hbox{$\sim$}}\ }
\def\gtap{\ \raise.3ex\hbox{$>$\kern-.75em\lower1ex\hbox{$\sim$}}\ }

\def\be{\begin{equation}}
\def\ee{\end{equation}}
\def\bea{\begin{eqnarray}}
\def\eea{\end{eqnarray}}

\newcommand{\Eref}[1]{Eq.~(\ref{#1})}
\newcommand{\gsim}{ \mathop{}_{\textstyle \sim}^{\textstyle >} }
\newcommand{\lsim}{ \mathop{}_{\textstyle \sim}^{\textstyle <} }
\newcommand{\vev}[1]{ \left\langle {#1} \right\rangle }

\newcommand{\gev}{{\rm GeV}}

\newcommand{\mev}{{\rm MeV}}

\newcommand{\fh}{\ensuremath{{\tilde h}}}
\newcommand{\fa}{\ensuremath{{\tilde a}}}
\newcommand{\fs}{\ensuremath{{\tilde s}}}
\newcommand{\h}{\ensuremath{{ h}}}
\newcommand{\A}{\ensuremath{{ a}}}
\newcommand{\s}{\ensuremath{{ s}}}

\newcommand{\hht}{\ensuremath{\tilde{h}}}
\newcommand{\st}{\ensuremath{\tilde{s}}}
\newcommand{\at}{\ensuremath{\tilde{a}}}

\title{Naturalness and Higgs Decays in the MSSM with a Singlet}
\author{Spencer Chang$^a$, Patrick J. Fox$^b$ and Neal Weiner$^a$\\
$^a$ Center for Cosmology and Particle Physics, Dept. of Physics, New York University,
New York, NY 10003 \\
$^b$ Theoretical Physics Group, Lawrence Berkeley National Laboratory, Berkeley, CA 94720\\
\email{sc123@cosmo.nyu.edu, PJFox@lbl.gov, nw32@nyu.edu}
}
\preprint{\today \\ }
\abstract{
The simplest extension of the supersymmetric standard model - the addition of one singlet
superfield - can have a profound
impact on the Higgs and its decays. We perform a general operator
analysis of this scenario, focusing on the phenomenologically distinct
scenarios that can arise, and not restricting the scope to the narrow
framework of the NMSSM. We reexamine decays to four b quarks and four
$\tau$'s, finding that they are still generally viable, but at the
edge of LEP limits. We find a broad set of Higgs decay modes, some
new, including those with four gluon final states, as well as more general 
six and eight parton final states. We
find the phenomenology of these scenarios is dramatically impacted by
operators typically ignored, specifically those arising from D-terms
in the hidden sector, and those arising from weak-scale colored
fields. In addition to sensitivity of $m_Z$, there are potential tunings of other aspects of the
spectrum. In spite of this, these
models can be very natural, with light stops and a Higgs as light as
82 GeV.  These scenarios motivate further analyses of LEP data 
as well as studies of the detection capabilities of future colliders to 
the new decay channels presented.
}
\keywords{Higgs Physics, Supersymmetry Phenomenology}
\begin{document} 

\section{Introduction}
Since 1934 when Fermi wrote down his theory of the weak interactions,
we have singled out the weak scale as an interesting scale for new
physics. Within the last two decades, we have finally reached this
energy scale at LEP and the Tevatron, and with the LHC we
should be able to probe this scale thoroughly.

The most pressing question at the weak scale is the origin of
electroweak symmetry breaking (EWSB). Within the
Glashow-Salam-Weinberg theory of weak interactions, it is broken by a
fundamental scalar doublet of $SU(2)$, the Higgs field. In this
framework, all precision quantities are calculable and agree with
present experimental limits.\footnote{We are of course neglecting dark
matter here, neutrino masses, the LSND anomaly and the NuTeV anomaly,
all of which are beyond the scope of this paper.} 

The major theoretical shortcoming of the standard model (SM) is the
question of the stability of the Higgs mass. Scalars in general
acquire quadratically divergent mass, suggesting that new physics
should cut off this divergence and appears near the weak scale.
Numerous solutions to this theoretical problem have been proposed,
most notably supersymmetry (SUSY).

\section{Naturalness in the MSSM}
In this paper we shall focus on supersymmetric solutions of the hierarchy problem. In these theories the introduction of superpartners cancels the quadratic divergences of the standard model with loops of opposite statistics particles, leaving only logarithmically divergent contributions to the Higgs mass, proportional to the SUSY breaking soft mass parameters.
 
Despite its excellent success in controlling divergences from very
high scales, in recent years SUSY has become far more constrained,
with superpartners pushed to higher scales, resulting in tunings which
are typically $O(4\%)$ in the minimal supersymmetric extension of the
standard model (MSSM).\footnote{For an excellent discussion of the
details of fine tunings in SUSY models, see \cite{Chacko:2005ra}.}
The reason for this is quite simple: the success of the LEP program has
pushed the lower limit on a SM Higgs boson above 114 \gev, a 
limit which  
applies in a large region of the MSSM parameter space. In the MSSM, the Higgs has a mass
which has a tree level upper limit of $m_Z = 91\ \gev$. To achieve a mass of 115 \gev, one
must invoke large radiative corrections due to (s)top loops. At one loop, the correction
to the Higgs mass is\cite{Martin:1997ns}
 \be
 \delta m_h^2 = \frac{3}{4 \pi^2}v^2 y_t^4 \sin^4 \beta \log(m_{\tilde t_r} m_{\tilde t_l}/m_t^2),
 \ee
 requiring a stop mass of roughly 500 \gev. Unfortunately, such a large stop mass also feeds into the soft mass squared of the Higgs,
 \be
 \delta m^2_{H_u} = -\frac{3 y_t^2}{8 \pi^2}(m_{\tilde t_{l}}^2 + m_{\tilde t_r}^2)\log(\Lambda/m_{\tilde t}),
 \ee
 where $\Lambda$ is the scale where the stop masses are generated. Even for low scale mass generation such as gauge mediation, where $\Lambda \gsim 16 \pi^2 m_{\tilde t}$, this results in a radiative correction to the up-type Higgs soft mass squared of $O((400\ \gev)^2)$. This must then be cancelled against a positive mass squared, for instance via a $\mu$-term, in order to achieve the appropriate vev, $v=174\ \gev$. This cancellation must be tuned at the level of $4 \%$ in order to achieve this. If the stops are heavier, or the mediation scale is higher, the tuning becomes worse.
 
The root of this problem is the small quartic of supersymmetric
models, which is fixed by the electroweak gauge couplings. There are
many proposals to enlarge the quartics, for instance by adding terms
in the superpotential
\cite{Fayet:1974pd,Nilles:1982dy,Frere:1983ag,Derendinger:1983bz,Duran
d:1989zs,Drees:1988fc,Ellis:1988er,Espinosa:1991gr}, with
non-decoupling D-term
quartics \cite{Batra:2003nj,Batra:2004vc, Maloney:2004rc}, or through strong dynamics
at an intermediate scale
\cite{Harnik:2003rs,Chang:2004db,Delgado:2005fq}. 
 
\subsection*{New Higgs Decays and Naturalness}
 An alternative approach is to evade the LEP limits indirectly.  For
instance, if the Higgs field decays in a non-standard way, LEP may not have been as sensitive to the decay. This is not
as trivial as it may appear, in particular because limits on invisible
Higgs decays are nearly as strong as those of b-quarks.  If these
non-standard decays are present in nature, there must be new states
lighter than the Higgs boson, which themselves decay into ordinary SM fields. The simplest possibility is the presence of an
additional singlet superfield to which the Higgs can decay, and which
then decays into SM particles, such as b-jets.
 In this vein, it has been recently argued that within the NMSSM,
where a singlet field $S$ acquires a vev to supply the $\mu$-term of
the Higgs sector, certain non-standard decays are possible, leading
to more natural theories \cite{Dermisek:2005ar}.
 
However, there is still a strong need for additional analyses, for many reasons.
\begin{itemize}
\item{New combined limits from LEP \cite{Sopczak:2005su,lepnote} exclude most
of the generic parameter space of these models. Typically, the decay $h\rightarrow 2a
\rightarrow 4b$ has been used to allow lighter Higgses. However, new
LEP combined analyses have basically raised the minimum Higgs mass for
this process to 110 \gev, nearly as strong as the limit on the
standard model Higgs. This raises the question of whether other
non-standard decays can occur generically, to which LEP analyses would
have been less sensitive. We will discuss these limits further in the
next section.}
\item{The NMSSM is not a fully general scenario. In the NMSSM, the
singlet acquires a vacuum expectation value to generate a $\mu$-term.
One typically requires that this be the true vacuum of the theory, and
additionally makes assumptions of the form of the theory in the
ultraviolet (UV). All of these things can distract from -- and are
beside the point of -- the basic phenomenological questions relating
to Higgs decays.}
\item{The sensitivity of $m_Z$ to UV parameters is not the only measure of naturalness in
these theories. Although a significant reason to consider these decays is precisely to
ameliorate this tuning, one often ends up with additional tunings in order to evade
experimental limits. Thus, while $m_Z$ may be relatively natural other elements of the
spectrum may be highly tuned, and necessary, to evade experimental limits.}
\item{The parameter space of new operators has not been fully explored. In the MSSM, the full set of soft breaking operators is considered. However, in adding a singlet to the theory, one can consider both the effects of D-terms in the supersymmetry breaking sector, and non-degenerate soft masses for the scalar and pseudoscalar components of the singlet, both of which can have significant phenomenological consequences.}
\item{The sensitivity of the decays of the pseudoscalar to the presence of new fields has essentially been ignored within the context of supersymmetric theories. Such fields can naturally induce the decay of the pseudoscalar to two glue jets, which have far weaker constraints.}
\end{itemize}

There are two important additional points to be made with regard to naturalness. The first is: what do we mean by tuning in the scenarios we are considering? Since we are interested in the low-energy phenomenological theory, it is impossible to quantify the sensitivity of $m_Z$ to the UV parameters. However, we know that the tuning of $m_Z$ arises in general due to the large values of the stop squark masses. Therefore, we shall use the stop masses as a proxy for this tuning. However, there is typically a tuning necessary to achieve the proper spectrum for the lighter scalar states, and this is usually the most severe tuning in the models. Therefore, we shall quote both stop masses, as well as the scalar mass spectrum tuning for every model considered.

The second point is: how natural is it to include a soft supersymmetry breaking operator, without an associated superpotential operator? In particular, how reasonable is it to include a trilinear scalar potential A-term operator without associated Yukawas? This question has been studied previously in the context of sterile neutrino masses in supersymmetry \cite{Arkani-Hamed:2000bq,Borzumati:2000mc}. In fact, if there is a SUSY breaking field which carries a charge, whether R-charge or Peccei-Quinn charge, it is quite natural for these operators to appear without an associated superpotential term. The converse, however, is not true. The presence of the superpotential term will radiatively generate the soft term at a minimum at the loop-suppressed level. Consequently, we will insist on technical naturalness, that these soft operators are present at least at this small level.

In lieu of these points, we will pursue a phenomenological
study of the effects of singlets on the decays and properties of Higgs
bosons. We will focus on decay modes that have not been considered
previously, including final states with 6 and 8 particles. We will often be studying situations when both the new scalar
and pseudoscalar states are lighter than the Higgs, and the Higgs is
light enough to have been produced through Higgs-strahlung at LEP,
although we will consider other scenarios. By also focusing on the
relevant parameters, we hope to clearly elucidate the effects of
mixing of the singlet with the Higgs, and the effect on the mass of
the Higgs boson.

Since this paper has both basic phenomenological points, as well as
points related to model building, we have attempted to lay out this
paper so that one who is interested principally in the phenomenology
can still learn the relevant points. Sections headed ``Model Building"
are independent, and the paper can largely be read without those.

The layout of the paper is as follows: in the next section, we will
review existing limits on Higgs decays, both for SM-like and
non-SM-like Higgs bosons.
In section
\ref{sec:singlets}, we will discuss the effects of singlets on Higgs
physics.  First, we show how non-standard decays into scalars can
arise and dominate the Higgs decay width.  We will also point out how
non-standard decays allow a larger mixing with the Higgs, this
mixing can raise the mass of the Higgs boson considerably without
resorting to radiative corrections. In its ``Model Building''
subsection, we present the relevant operators that induce the Higgs
mixing and decays.  In section \ref{sec:scenarios}, we will sketch out
the spectra and non-standard decay modes which are consistent with
existing LEP
bounds, and lead to a more natural parameter space of the
theory. In
its ``Model Building'' subsections,  we discuss model realizations of
these different spectra and decay scenarios. 
In doing so, we can assess more simply the degree of tuning required
to satisfy the experimental constraints.  In section
\ref{sec:benchsum}, we summarize the presented benchmark points and
their relevant phenomenology for future collider experiments.      
In section \ref{sec:conclusions}, we discuss additional directions
worthy of further investigation and conclude.  In particular, we suggest some new analyses on 
LEP data which might be useful when considering these scenarios.  Finally,
in appendix \ref{sec:trilinears} some calculational results for scalar
trilinears are summarized.
 
\section{Summary of LEP limits on Higgs}
Most particle physicists are familiar with the LEP2 95\% CL limit on
the SM Higgs of $m_h \geq 114.4 \;\gev$, but 
LEP has also produced a multitude of other limits on Higgs physics.  

The limits that are most applicable to this work are the 95\% CL
limits on the so called $\xi^2$ parameters (often also referred to as
$C^2$).  We are analyzing a 
SM-like Higgs, which means that Higgs-strahlung production of the Higgs
is close to the SM rate.\footnote{More generally, we will usually be
taking the decoupling limit, where the lightest Higgs state is
produced by Higgs-strahlung, but
not in associated production with the CP-odd $A^0$.  Thus, constraints
from Higgs-strahlung are the main concern in these scenarios.}  
Then depending on the assumed process of Higgs decay $h \to X$, the
defined limits are on 
$\xi^2 \equiv \frac{\sigma(e^+e^- \to hZ)}{\sigma(e^+e^- \to
hZ)_{SM}}BR(h \to X)$.  We now list the different limits that
LEP has analyzed, based on Higgs-strahlung and the given Higgs decay
process (note: all stated mass limits assume $\xi^2 =1$):

\begin{itemize}
\item{Model Independent Decays: This is the most conservative limit on the Higgs boson.  It assumes that the Higgs is 
produced with a Z boson and looks for electrons and muons that reconstruct to a Z mass, while the Higgs decay process is unconstrained
(by theory or the event analysis).  
The only study of this sort is done by OPAL giving a limit of $m_h \geq 82 \;\gev$, see Fig.~11 of \cite{Abbiendi:2002qp}. 
Unfortunately, no other collaboration
has released such an analysis.  }
\item{Standard Model Higgs: LEP-wide 
limits on the SM Higgs are given in Fig.~10 of \cite{Barate:2003sz},
requiring $m_h \geq 114.4 \,\gev$.  This study also includes the strongest limits on $h
\to b\bar{b},\tau \bar{\tau}$ rates.}
\item{Invisible Decays: In this analysis, the
Higgs is assumed to decay into stable (on collider length scales)
neutral particles.  The implication is that non-standard Higgs decays
have to primarily decay into visible particles.  
Both L3 and DELPHI have performed such an analysis \cite{Achard:2004cf,Abdallah:2003ry}, but the most stringent constraints are from an older preliminary LEP-wide analysis that has a limit $m_h \geq 114 \,\gev$, see Fig.~4 in \cite{unknown:2001xz}.}
\item{2 Photon Decays (aka Fermiophobic):  Fermiophobic typically means a Higgs with standard couplings to gauge bosons, but suppressed couplings to fermions, allowing decays into $WW^*$, $ZZ^*$ as well as photons. If all decay modes are open, there is a limit of $m_h
\geq 109.7 \ \gev$, while decays exclusively to two photons have a limit of $m_h
\geq 117 \;\gev$. See Fig.~2 of \cite{Rosca:2002me}, which is the LEP-wide analysis.}
\item{2 parton hadronic states (aka Flavor-independent): In this type of analysis, any two jet decays of the Higgs are allowed. 
The analyses use the jets that are least sensitive to the
candidate Higgs mass and details of the Z decay.  Each LEP collaboration has done a study
\cite{Abdallah:2005cv,Achard:2003ty,Abbiendi:2003gd,Heister:2002cg}.  However, the
strongest constraint is the preliminary LEP-wide analysis of $m_h \geq 113 \;\gev$, see
Fig.~2 of \cite{:2001yb}.}
\item{Cascade Decays:
These are the constraints that are most relevant for the present
study, where cascade decays mean that the Higgs decays into
two scalars $\phi$ and those scalars decay into $X$ (i.e. $h \to 2 \phi \to 2X$).  
OPAL \cite{Abbiendi:2004ww} and
DELPHI \cite{Abdallah:2004wy} looked at b decays ($X \equiv b\bar{b}$), see their Fig.~12's and a new 
LEP-wide analysis \cite{Sopczak:2005su,lepnote} has constrained both b and $\tau$ decays, with the exclusion plots given on page 8 in the first reference. For
$h\rightarrow 2\phi \rightarrow 4b$ the limits are now 110 \gev\ for a
Higgs produced with SM strength.  For other intermediate scalar decays,  $\phi \to 2g, c\bar{c}, \tau\bar{\tau}$, the best model independent exclusions are from OPAL's analysis when the mass of the scalar is below $b\bar{b}$ threshold, which is given
in Fig.~7 of \cite{Abbiendi:2002in} (note: the analysis is restricted for Higgs masses in the
range $45-86 \;\gev$).}  
\end{itemize}

\section{Singlets and the MSSM}
\label{sec:singlets}
As we have previously said, the simplest extension of the MSSM is to
add a SM singlet superfield $S$, containing a CP-even scalar $s$, and
a CP-odd pseudoscalar $a$.\footnote{More precisely, the CP-even properties of $a$ are not determined until its interactions and mixings are given.  
For example, it can pick up a CP transformation if it mixes with a scalar with fixed CP properties (the CP-even $h^0,H^0$ or CP-odd $A^0$) or couples to fermions/gauge bosons in a certain CP fashion.  On the other hand, in all cases $s$ mixes with $h^0$ and thus is always CP-even.}
The dominant phenomenological effects of this
new field are:
\begin{itemize}
\item{New decays for the Higgs boson. If one or both of the new
states are lighter than  half the Higgs mass, decays $h\to 2s$
and $h\to 2a$ are possible, followed by $s\to X$ or $a\to X$.}
\item{Light states can mix with the Higgs boson. If a light state
mixes with the Higgs boson, it can push the mass of the Higgs boson up
without large radiative corrections. This is at the cost of having a
new light state which can be produced through brehmstrahlung off a Z,
for which there are stringent constraints.}
\end{itemize}
 
We first show that non-standard decays can easily dominate over SM decays.
In the case of cascade decays, for a Higgs produced with SM strength
through Higgs-strahlung, the new process must have a width at
least\footnote{This condition is not sufficient if the scalar $a$ decays into b quarks, as will be discussed later in section \ref{sec:mixing}.} $\Gamma_{h\to 2 a}\gsim 4 \times \Gamma_{h\to 2 b}$. With a term in
the Lagrangian
\be
\frac{c v}{\sqrt{2}}h a^2
\ee
the width to two scalars is \cite{Dobrescu:2000jt}
\be
\Gamma_{h\to 2 a}=\frac{c^2 v^2}{16 \pi\, m_h}\left(1- \frac{4 m_a^2}{m_h^2} \right)^{1/2}
\ee
which one can compare with the dominant width of the Higgs to b quarks (in this mass range)
\be
\Gamma_{h\to b\bar{b}} = \frac{N_c m_b^2}{16\pi v^2}m_h \left(1-\frac{4m_b^2}{m_h^2} \right)^{3/2} \sim .003\; \gev \left(\frac{m_h}{100\; \gev}\right).
\ee
Taking into account higher order effects, a more accurate approximation for the Higgs decay width to SM particles in the mass range of interest
is (using HDECAY \cite{Djouadi:1997yw})
\be
\Gamma_{h\to SM}=(2.34\times 10^{-5} m_h+.206 \,\mev)
\ee
For a 100 \gev\ Higgs, in order for scalar decays to sufficiently dominate, one must have
\be
\frac{c v}{\sqrt{2}} \gsim 5\; \gev
\ee
The required size of the trilinear indicates that non-standard cascade decays can easily dominate over standard Higgs decays.
 
One can also study the effects of mixing with a singlet on the Higgs mass in a model independent fashion.  In the presence of mixing, the mass eigenstates \fs\ and \fh\ will be related to the interaction eigenstates through a mixing matrix
\be
\label{eqn:rot}
\pmatrix{ \fs \cr \fh} = \pmatrix{\cos\theta & -\sin \theta \cr \sin
\theta & \cos \theta} \pmatrix{ s \cr h}
\ee
Similarly for the mixing in the CP-odd sector,
the mass eigenstates \at\ and $\tilde{A}$ are related to the
interaction eigenstates by
\be
\pmatrix{\at \cr \tilde{A}}=\pmatrix{\cos\phi &- \sin\phi \cr
\sin\phi  & \cos\phi}\pmatrix{a \cr A^0}
\ee
If the MSSM mass of the Higgs is
$m_{mssm}^{2}$, then the mass eigenvalue after mixing with the lighter
singlet is
\be
m_\fh^2 = \frac{m_{mssm}^{ 2}- m_{\fs}^2 \sin^2\theta}{\cos^2\theta},
\ee
where $m_{\fs}$ is the mass of the mass eigenstate \fs.  This increase in mass is only through mixing and not through radiative corrections.

In the subsection below, we discuss some of the operators which can be used in building models with certain phenomenologies.  Those readers who are only interested in the descriptions of the phenomenology can proceed to section \ref{sec:scenarios}.

\subsection*{Model Building: New Operators with Singlets}
The introduction of a singlet superfield allows us to introduce a number of new operators in the theory.
\TABLE[t]{
\begin{tabular}{|c|c|l|}\hline
\ensuremath{\delta_s^2} & $X X^\dagger (S + S^\dagger)^2_D  $& mass for \s\\ 
\hline
\ensuremath{\delta_a^2} & $X X^\dagger (S - S^\dagger)^2_D $ & mass for \A \\ 
\hline  \hline
\ensuremath{m_D} & $(W_\alpha' W^\alpha S)_F $ & singlino (Dirac) mass, mixes \s\
and \h \\ && 
        allows $h\to 2s$ decays, allows Z \s\ production \\
\hline
\ensuremath{\lambda_h} & $(S H_u H_d)_F$  & (with mu term) mixes \s\ and \h , \s\ and
\A\ masses, \\ && singlino (Dirac) mass, allows $h\to 2s$, $h\to 2a$
decays   \\ 
\hline
\ensuremath{A_s} & $(X S^3)_F\ +\ h.c.$    & Real: (with $\vev{s}$) mass for \s\,
$s\to 2a$ decays, 3 \s\ coupling\\ && Imaginary: $a\to 2s$ decays, 3
\A\ coupling  \\ 
\hline
\ensuremath{A_h} & $(XS H_u  H_d)_F\ +\ h.c.$    & mixes \s\ with  $H$, $a$ with
$A^0$, $a$ decays (due to mixing) \\ &&
with mixing, allows $h\to  2s,\ 2a$ decays (esp. large $\tan\beta$)
\\ 
\hline
\ensuremath{m_{CP}^2} & $X X^\dagger (S + S^\dagger) (S - S^\dagger)_D$   &  Mixes \s\ and \A,
allows $a$ decays (due to mixing)   \\ 
\hline
$M_Q^{-1}$ & $(S Q \bar Q+M_Q Q \bar Q)_F$  &  Allows $s$ and $a$ decays to
gluons and photons   \\ 
\hline
\end{tabular}
\caption{Operators involving the Singlet\label{tb:ops}}}
These operators have the effects of introducing into the theory
masses, mixing, and decays of the states.  The operators are listed in
Table \ref{tb:ops}. We will describe here the effects of these
operators when they appear singly. Clearly, the effects of the
operators do not necessarily add linearly, particularly when vevs turn
on, but it is useful to understand their effects individually. In many
scenarios this is sufficient to understand the phenomenology and we
will point out the situations where there are interference effects
from multiple operators.

\subsection*{\ensuremath{\delta_s^2}, \ensuremath{\delta_a^2} -- Scalar and Pseudoscalar Soft Masses}
The operators arising from K\"ahler potential terms, \ensuremath{\delta_s^2} and \ensuremath{\delta_a^2},
determine the spectrum of the theory by adding mass terms to the
Lagrangian
\be
{\cal L}_{\ensuremath{\delta_s^2}} = -\frac{\delta_s^2}{2} s^2 \hskip 0.6 in {\cal L}_{\ensuremath{\delta_a^2}} = -\frac{\delta_a^2}{2} a^2.
\ee
Most previous analyses include equal soft masses for the scalar and
pseudoscalar, although there is often no a priori symmetry reason to
do so. All other scalar fields in the MSSM carry a gauged U(1), forcing a
degeneracy which is not necessary for singlets. Because our interests
are phenomenological, we allow different masses, and thus more 
interesting spectra. In particular, this allows spectra of the form $m_h > 2 m_s >
4 m_a$, leading to events with many final state jets, b-quarks and
photons, produced through cascades.  

The tuning of these theories with singlets, when $m_h< 114 \;\gev$, is usually encoded in how tuned these additional masses must be in order to achieve a proper spectrum, and not in the sensitivity of $m_Z$ or the masses of the stops. When we refer to these operators, a value of \ensuremath{\delta_s^2} or \ensuremath{\delta_a^2} of $\pm (20 \gev)^2$ refers to a contribution of $\pm (20 \gev)^2$ to the scalar or pseudoscalar mass squared.

\subsection*{\ensuremath{m_D} -- Supersoft Operator}
\ensuremath{m_D} is the so-called ``supersoft'' operator \cite{Fox:2002bu} and arises from a SUSY breaking 
spurion in the superpotential. This operator is particularly interesting, and often neglected. 
If the hidden sector has a $U(1)'$ which acquires a D-term expectation value, it has no effect in the MSSM, in that all soft operators can be generated by F-term breaking. 
On the other hand, in the presence of a singlet, D-term breaking
becomes important through the introduction of this new
operator\cite{Carpenter:2005tz}. The Lagrangian terms generated are
\be
{\cal L} = \int d^2 \theta \frac{W^\prime_{\alpha}}{M} W_Y^\alpha S+
h.c.
\rightarrow -\frac{1}{2}(m_D s+ D_Y)^2
\ee
where $D_Y$ is the usual hypercharge D-term, $D_Y = \sum_i g_Y q_i
\phi^*_i \phi_i$, and $\frac{W^\prime_\alpha}{M}=\theta_\alpha m_D$.
Although it is supersymmetry breaking, it does not feed into the RG flow of any other soft operators, hence the term
supersoft. The principal effects are to produce a mass for $s$ (but
not $a$), and a coupling $s h h^*$ which is proportional to the new
contribution to the $s$ mass. It also generates a Dirac mass for the
Bino with $\psi_S$ equal to $m_D/2$. When we refer to a value of $\ensuremath{m_D}$
of 20 \gev, this implies a contribution to the scalar mass squared of $(20
\gev)^2$, and the corresponding trilinear term.

This operator directly affects the physics of electroweak symmetry breaking and the properties of the Higgs. However, it is simple to perform the same minimization of the Higgs potential in the presence of this term. Let us consider the usual MSSM neutral Higgs potential with this contribution.
\bea
\nonumber
V = &&(|\mu|^2+m_{H_u}^2) H_u^2 + (|\mu|^2+m_{H_d}^2) H_d^2
-(b H_u H_d + h.c.) \\&& +\frac{g^2}{8}(H_u^* H_u -
H_d^* H_d)^2\\ && +\frac{1}{2}\left[m_D s + \frac{g_Y}{2}(H_u^* H_u -
H_d^* H_d)\right]^2+\frac{\Delta^2_s}{2} s^2
\nonumber
\eea
where $\Delta_s^2$ encodes other contributions to the \s\ mass, such as from \ensuremath{\delta_s^2}.

When electroweak symmetry is broken, there is a linear term for \s, and so we shift \s\ by an amount
\be
s\rightarrow s+ \frac{g_Y m_D v^2 \cos 2 \beta}{2(m_D^2 + \Delta_s^2)}
\ee
The Higgs potential (neglecting \s\ which has already been shifted to its appropriate minimum) now reads
\bea
V = &&\left[|\mu|^2+m_{H_u}^2+\frac{g_Y^2 m_D^2 v^2 \cos 2 \beta}{4(m_D^2 + \Delta_s^2 )}\right] H_u^2 + \left[|\mu|^2+m_{H_d}^2-\frac{g_Y^2 m_D^2 v^2 \cos 2 \beta}{4 (m_D^2 + \Delta_s^2 )}\right] H_d^2  \\&& -(b H_u H_d +
h.c.) + \frac{(g^2 + g_Y^2)}{8}(H_u^* H_u - H_d^* H_d)^2
\nonumber
\eea
So we see the presence of the $s$ vev amounts to a redefinition of the values of $m_{H_u}^2$ and $m_{H_d}^2$. Thus, we can perform the usual minimization and diagonalization of the MSSM Higgs fields. The new trilinear couplings induce a mixing between the \s\ and the other Higgses, which is encoded in the mass matrix (in the $(s\ h\ H)$ basis)
\be
\pmatrix{m_D^2 + \Delta_s^2 & \frac{g_Y m_D v  s_{\alpha+\beta}}{ \sqrt{2}} & -\frac{g_Y m_D v  c_{\alpha+\beta} }{\sqrt{2}} \cr
\frac{g_Y m_D v  s_{\alpha+\beta}}{\sqrt{2}} & m_h^2 &0 \cr
-\frac{g_Y m_D v  c_{\alpha+\beta} }{\sqrt{2}} &  0 &m_H^2}
\ee
Here, as we dial up $m_D^2$, which is always positive, we increase the mixing. The values of $m_h^2$ and $m_H^2$ are the usual ones from the MSSM, in particular with $m_h^2$ bounded at tree level to be below $m_Z^2$.

Since $s_{\alpha+\beta}\equiv \sin(\alpha+\beta) > \cos(\alpha+\beta)\equiv c_{\alpha+\beta}$ generally, and
$m_H^2>m_h^2$, we can just focus on the $2\times 2$ mixing submatrix
between $s$ and $h$.  The mass eigenstates are related to the
interaction eigenstates as given in (\ref{eqn:rot}).
We are principally interested in the trilinear terms, especially the $\fs^2 \fh$ term,
which allows $\fh\rightarrow 2 \fs$ decays.
There are two contributions to this term, one arising from the new
supersoft operator, once we have mixed the CP-even states, and one
from the usual D-term trilinears, once we have mixed. The coefficients
of these terms are given in the appendix.

\subsection*{$\lambda_h \mathbf{S H_u H_d}$}
\ensuremath{\lambda_h} is a commonly studied superpotential operator, as it induces an additional Higgs
quartic at small $\tan \beta$ which can raise the tree level Higgs mass
\cite{Ellis:1988er,Drees:1988fc,Espinosa:1991gr}.
After electroweak symmetry breaking, it generates masses for \A\ and
\s\ of size $\lambda_h v$. 
With these contributions the potential is given by
\bea
\nonumber
V = &&(|\mu|^2+m_{H_u}^2) H_u^2 + (|\mu|^2+m_{H_d}^2) H_d^2
-(b H_u H_d + h.c.) +\\&& \frac{(g^2 + g_Y^2)}{8}(H_u^* H_u -
H_d^* H_d)^2+\lambda_h^2 (H_u H_d) (H_u H_d)^*\\&&
+ \frac{\lambda_h^2}{2}(s^2+a^2)(H_u^*H_u+H^*_dH_d)+\sqrt{2}\lambda_h \mu
s(H_u^*H_u+H^*_dH_d)+\frac{\delta^2_s}{2} s^2 +\frac{\delta^2_a}{2}
a^2
\nonumber
\eea
Note the explicit $\mu$-term for the Higgses (and the resulting
trilinears), this is to be compared to the NMSSM where the vev of $S$
gives the $\mu$-term. 
However, absent any additional mass terms for $S$, i.e. $\delta_s$, we
can simply shift
$S\to S-\sqrt{2}\mu/\lambda_h$, removing the entire $\mu$ term.
In the presence of of additional soft masses for $s$ there is an s-vev of,
\be
\vev{s}=-\frac{\sqrt{2}\lambda_h\mu
v^2}{\delta_s^2+\lambda_h^2 v^2}.
\ee
 The effect of this is to replace $\mu$ by an effective  $\mu$,
which we denote $\tilde{\mu}$, where
 \be
 \tilde \mu = \mu \left(\frac{\delta_s^2}{\delta_s^2+\lambda_h^2 v^2} \right).
 \ee

Through the new quartic and trilinear terms there are mixings
between \s\ and the MSSM Higgses, giving a mass matrix
\be
\pmatrix{\lambda_h^2 v^2 +\delta_s^2& - 2\lambda_h v\tilde\mu
s_{\alpha-\beta} & 2\lambda_h v \tilde\mu c_{\alpha-\beta} \cr
 - 2\lambda_h v\tilde \mu s_{\alpha-\beta} & m_h^2 & 0 \cr
 2\lambda_h v \tilde \mu c_{\alpha-\beta} & 0 & m_H^2}
 \ee
Notice, however, that in the absence of additional soft masses for
\s\ ($\delta_s^2 = 0$) the mixing vanishes, and increases as we
deviate from zero.
The presence of \s-\h\ mixing allows $\fs \rightarrow 2a$
decays.  It is possible for $\fh\rightarrow 2\fs, 2a$ decays to occur with
equal amplitude, for instance at small mixing. However, it is
also possible to arrange for \ensuremath{\lambda_h} to give only one dominate decay of
$\fh$.  Finally, remember that 
the presence of \ensuremath{\lambda_h} naturally requires the presence of \ensuremath{A_h},
at least at a loop-suppressed level.

\subsection*{$A_s \mathbf{S^3} + c.c.$}
Another possible operator to add is a soft trilinear for $S$ in the
potential, $A_s S^3 + c.c.$.  Alone this has no effect on the Higgs,
although it does give equal but opposite contributions to $s$ and $a$ soft masses with an
$S$ vev and also induces $s\rightarrow 2a$ decays. However, if there is another
source of $s-h$ mixing, \ensuremath{A_s} can generate $\fh\rightarrow 2\fs$ and
$\fh\rightarrow 2a$ decays as well as the original $\fs\rightarrow 2a$. 
The effective potential with this contribution is,
\bea
V = &&(|\mu|^2+m_{H_u}^2) H_u^2 + (|\mu|^2+m_{H_d}^2) H_d^2
-(b H_u H_d + h.c.) +\\&& \frac{(g^2 + g_Y^2)}{8}(H_u^* H_u -
H_d^* H_d)^2+ \frac{A_s}{\sqrt{2}}(s^3-3s a^2)
+\frac{\delta^2_s}{2} s^2 +\frac{\delta^2_a}{2} a^2
\nonumber
\eea

Since this operator does little of interest by itself it must be
considered in tandem with some other operator that produces $s-h$
mixing and an s-vev.  One combination of note is \ensuremath{m_D} and \ensuremath{A_s}.  We can
analyze the general case under the simplifying assumption, as above,
that the mixing involves mainly $s$ and $h$ and not $H$, so the
mass and interaction bases are related as in \Eref{eqn:rot}.

The relevant trilinears are given in the appendix. As in the case of \ensuremath{\lambda_h}, at small mixing $\fh\rightarrow 2\fs$
and $\fh \rightarrow 2a$ have equal amplitudes.
The size of the s-vev $\vev{s}$, generated by the mixing operator will be
corrected by \ensuremath{A_s} but if $A_s\vev{s}^2\ll m_s^2 \vev{s}$, as will often
be the case, its
effects can be ignored.  

\subsection*{$A_h \mathbf{S H_u H_d} + c.c.$}
The potential with \ensuremath{A_h} included is,
\bea
V = &&(|\mu|^2+m_{H_u}^2) H_u^2 + (|\mu|^2+m_{H_d}^2) H_d^2
-(b H_u H_d + h.c.) +\\&& \frac{(g^2 + g_Y^2)}{8}(H_u^* H_u -
H_d^* H_d)^2+ ( A_h S H_u H_d + c.c)
+\frac{\delta^2_s}{2} s^2 +\frac{\delta^2_a}{2}
a^2 \nonumber
\eea
After EWSB this generates a vev for s,
\be
\vev{s}=-\frac{A_h v^2}{\sqrt{2}\delta_s^2} \sin 2\beta
\ee
This gives a correction to the b-term for the MSSM Higgses, but still
allows the usual diagonalization, leading to a mixing matrix of the
form, 
\be
\pmatrix{
\delta_s^2 & A_h v c_{\alpha +\beta} & A_h v s_{\alpha +\beta} \cr
 A_h v c_{\alpha +\beta} & m_h^2 & 0 \cr
A_h v s_{\alpha +\beta} & 0 & m_H^2
}
\ee
Unlike the previously discussed operators, \ensuremath{A_h} mixes the CP odd
component of $S$ with $A^0$.  The mixing matrix, in the basis ($a$ 
$A^0$), is
\be
\pmatrix{
m_{a}^2 & -A_h v \cr
-A_h v & m_{A}^2 
}
\ee
The mass eigenstates are related to the interaction eigenstates as
given in (\ref{eq:amix}).
This mixing of $a$ with $A^0$ means that $a$ can decay to 2 b-quarks
or 2 $\tau$'s if its mass is below about 10 GeV.  With no additional
terms in the Lagrangian, e.g. \ensuremath{M_Q^{-1},m_{CP}^2=0}, this is the only decay path of $a$. The relevant trilinears are given in the appendix.
 
\subsection*{$m^2_{CP}\, \mathbf{s\,a}$ -- CP Mixing Mass}
The operator \ensuremath{m_{CP}^2} is a CP mixing mass which mixes the would-be pseudoscalar
$a$ with the would-be scalar $s$. This operator can be important in
allowing decays of the $a$ when \ensuremath{M_Q^{-1}} and \ensuremath{A_h} are zero and can be
particularly important when decays $h\to 2s$ would dominate the Higgs
decay, but are kinematically forbidden. In this case, mixing due to \ensuremath{m_{CP}^2}
can make $h\to sa$ the dominant decay mode.

We should point out that this operator does not, {\it by itself}, violate CP. As mentioned earlier, without any other interactions, there is nothing to prevent us from assigning CP-even transformation properties to both $a$ and $s$. This remains true with the inclusion of $A_s$ or the supersoft operator $m_D$ in the potential. Only when $a$ couples to fermions or gauge bosons as a pseudoscalar, or mixes with the $A^0$ is a CP-odd property forced upon it. The scenarios which we will study will not have such a mixing and hence no actual CP violation is introduced into the theory, making such a term safe from edm searches, for example.  

One requires either \ensuremath{m_D} or \ensuremath{\lambda_h} to induce Higgs decays (through direct couplings or mixings). In the case that \ensuremath{\lambda_h} is present, $h\to 2a$ is already significant, so $h \to as$ typically cannot compete. However, in the \ensuremath{m_D} case, where $a$ has no couplings, such a scenario is viable. We thus consider the mass matrix (in the ($a\ s\ h)$) basis)
\be
\pmatrix{
\delta_a^2 & m^2_{CP} & 0 \cr
m^2_{CP} & m_D^2 + \delta_s^2 &  \frac{g_Y m_D v  s_{\alpha+\beta}}{ \sqrt{2}} \cr
0 &  \frac{g_Y m_D v  s_{\alpha+\beta}}{ \sqrt{2}} & m_h^2
}
\ee
Because the mixing will allow the $a$ to be produced through
$a$-strahlung from the Z, LEP limits become very severe. Ultimately,
this implies that \ensuremath{m_{CP}^2} must be somewhat small, and for our purposes here, we will treat it as a
perturbation. Ultimately, it will be necessary to calculate mixings precisely, but for estimates and intuition, it is easiest to study in the perturbative limit. We can perform a rotation in the $sh$ sector by an
angle $\theta$, which may be large. This leaves the following matrix
\be
\pmatrix{
m_a^2 & m^2_{CP} \cos \theta & m^2_{CP} \sin \theta \cr
m^2_{CP} \cos \theta & m_{\tilde s}^2 & 0 \cr
m^2_{CP} \sin \theta & 0 & m_{\tilde h}^2 
}
\ee 
The mass matrix can then be diagonalized by a series of rotations
\be
U=R(\theta_{sh}) R(\delta_{a s}) R(\delta_{ah})
\ee
where $\theta_{sh}=\theta$ diagonalizes the $s h$ mass matrix, and the $R(\delta_i)$ are perturbative rotations where $\sin \delta_{as} = m^2_{CP} \cos\theta/(m_{\tilde s}^2 - m_{a}^2)$ and $\sin \delta_{ah} = m^2_{CP} \sin\theta/(m_{\tilde h}^2 - m_{a}^2)$, where $m_{\tilde s,\tilde h}^2$ are the $\tilde s \tilde s$ and $\tilde h \tilde h$ entries after the $\theta$ rotation.

The relevant mixing angle which enters into $\tilde a$-strahlung is 
\be
\label{eq:amix}
\sin \theta_{ah}=\sin \theta \cos \delta_{ah} \sin \delta_{as} -\cos \theta \sin \delta_{ah} \simeq m^2_{CP} \sin\theta \cos \theta \left( (m_{\tilde s}^2 - m_a^2)^{-1}-(m_{\tilde h}^2 - m_a^2)^{-1} \right).
\ee

\subsection*{\ensuremath{M_Q^{-1}} -- Fermiophobic Decay Operator}
If the singlet superfield couples to new vector matter, there are loop induced decays of the scalars into gauge bosons.  
The presence of the new fields corrects the beta function of the SM gauge groups, with a superpotential coupling
\be
\lambda_Q S Q \bar Q + M_Q Q \bar Q
\ee
the 
gauge kinetic term becomes
\be
\int d^2\theta\left( \frac{1}{2g^2}+\frac{b_i}{16\pi^2} \log(m_Q +
\lambda_Q S)\right) W_\alpha W^\alpha \simeq \left( \frac{1}{2g^2}+\frac{b_i}{16\pi^2} M_Q^{-1} 
S\right) W_\alpha W^\alpha
\ee
where $W_\alpha$ is the gauge field strength, and $M_Q^{-1} = \lambda m_Q^{-1}$.
This results in a coupling between $s$ and two gauge fields,
as well as $a$ and two gauge fields which explains why we refer to this as a fermiophobic decay operator. Because $s$ generally mixes with the Higgs, it is difficult for a loop suppressed decay to compete. However, $a$, which mixes with the often heavy $A^0$, and often through the loop
suppressed operator $\ensuremath{A_h}$, can have its dominant decay mode through
this operator. Expanding out the expression above, we find a term in
the Lagrangian
\be
{\cal L}\supset c (a\, F^a \tilde{F}^a) \equiv c (a\, 
\epsilon^{\mu\nu\rho\sigma} F^a_{\mu\nu} F^a_{\rho\sigma})/2
\ee
This induces $a$ decays into photons or gluons, with the decay width
\begin{equation}
\Gamma_c = \frac{c^2}{4 \pi} m_a^3 N_D
\end{equation}
where $N_D$ is the number of ``colors'' in the final state (i.e. 1 for 
photons and 8 for gluons).

With the convention above, 
\begin{equation}
c = \frac{\sqrt{2} M_Q^{-1}}{64 \pi^2 } b_i g_i^2 = \frac{\sqrt{2} 
M_Q^{-1}}{16 \pi} b_i \alpha_i
\end{equation}   
where $i$ labels the gauge group and $b_i$ is it's beta function  
where $\frac{dg_i}{d \log{\mu}} = b_i 
\frac{g_i^3}{16\pi^2}$.
In the specific case of a $5+\bar{5}$ and higgsinos 
\begin{equation}
\begin{array}{c|c|c|c}
& D,D^c&L,L^c&\tilde{H}_u,\tilde{H}_d \\
\hline 
b_3 &1&0&0\\
b_Q &2/3&2&4/3 \\
\end{array}  
\end{equation}
We can compare the decays into gluons and those into b quarks through
mixing,
\begin{equation}
\Gamma_{glue}/\Gamma_{b\bar{b}} = \frac{(N_c^2-1) c^2 m_a^3/(4 
\pi)}{(m_b/v)^2 N_c 
(\sin\phi \tan \beta)^2 \sqrt{1-4m_b^2/m_a^2} (m_a/16\pi)}.
\end{equation}
which for the particular case of a $D,D^c$, becomes
\begin{equation}
\Gamma_{glue}/\Gamma_{b\bar{b}} = \frac{1}{4 \pi^2}\frac{\alpha_3^2
}{N_c (\sin\phi \tan 
\beta)^2 \sqrt{1-4m_b^2/m_a^2}}\left(\frac{m_a v}{m_b M_Q}\right)^2
\end{equation}
For decays to glue to dominate over b's, one requires small mixing
angles in the CP-odd sector, which can be achieved simply with loop
suppressed \ensuremath{A_h}, or a heavy $A^0$. Alternatively, one can have lighter
$D,D^c$ squarks, for example, with the values $\alpha_3 = .12, m_a =
40 \gev, M_Q
= 330 \gev$ one requires $(\sin\phi \tan\beta) \lsim 
.05$ to get comparable rates.

When $M_Q^{-1}$ induced decays for \fa\ dominate, we can get an estimate on the branching ratios into gluons and photons in the case of $S$ coupling to a $SU(5)$ complete multiplet.  For this, the branching ratios are $(2g, 2\gamma) = (.99, 3.8 \times 10^{-3})$.  If the Higgs dominantly decays into 2\fa, this gives branching ratios $\fh \to 2\fa \to (4g, 2g\, 2\gamma, 4\gamma)$ of $(.99,\, 7.6 \times 10^{-3},\, 1.5 \times 10^{-5})$.  The channels with photons may be enhanced by including \ensuremath{\lambda_h} which only induces decays into photons, via Higgsino loops.
As a final comment on this operator, we point out that if this is the only allowed \fa\ decay, as $M_Q$ increases, \fa\ will first decay with a visible displaced vertex and eventually will decay outside of the detector.  Both of these possibilities are highly constrained by LEP searches if the Higgs can be produced, ruling out such a Higgs that cascade decays into two such \fa\ scalars.\footnote{Note added: It has recently been emphasized that similar Higgs decays with highly displaced vertices could well be visible at Tevatron and LHC, enhancing its detection prospects \cite{Strassler:2006ri}.}  Therefore, we only consider values of this operator where \fa\ decays promptly.       
 
\subsection*{Other Operators}
Note that we ignore here the effects of a $\mu$-term for the $S$
field, and a superpotential $S^3$ term. Since we are mostly concerned
with the scalar sector of the theory, and aren't concerning ourselves
with the cosmological implications at this time, the mass of the
singlino is irrelevant for our purposes. The only effect of $\mu_S$ is
then to generate an $S^* H_u H_d$ term in the potential in a cross
term with \ensuremath{\lambda_h}. This has the same phenomenological effect as \ensuremath{A_h}, so
including $\mu_S$ should not change the basic outcome of our study.
Including $S^3$ in the superpotential can induce invisible decays of
the Higgs to a singlino, but these are highly constrained and not
interesting for our purposes. It can also induce a term $H_u H_d S^{*2}$
in the potential, but this acts simply like \ensuremath{A_h} in the presence of an
$s$-vev, and like \ensuremath{\lambda_h} in the presence of an $h$-vev in allowing
$h\rightarrow 2a, 2s$ decays. Hence, we do not consider either
operator here.

\section{Scenarios}
\label{sec:scenarios}
Although we have expanded the MSSM only by a single superfield, there is a remarkably large number of new scenarios of Higgs decays which arise, with large variations in the number of particles in the final state, as well as types of particles in the final state.

We can group the scenarios, in increasing order of complexity, into three categories:
\begin{itemize}
\item{The Higgs is mixed heavy. In the presence of large mixing, the Higgs mass can be increased significantly, in spite of small radiative corrections. In this scenario, the Higgs is at least 114 \gev\ in the case of SM-like decays, or 106 \gev\ in the case of cascade decays to 4 b quarks.}
\item{The Higgs decay is dominated by a single stage decay. Here $h\to 2a \to 2X$ or $h\to
2s \to 2X$, where $X$ is composed of a pair of standard model fields. As above, if $X=2b$,
one requires $m_h\gtap 106 \gev$, however, if $X= 2 \tau$ or $X=2 g$, the Higgs can be
considerably lighter. }
\item{The Higgs decay is dominated by a two-stage cascade. That is
$h\to 2s \to 4 a \to 4X$ or $h \to as \to 3a \to 3X$. Such two stage cascades generally do not occur if one
restricts oneself to the scenario of the NMSSM. Here $X=2 \tau, 2 b, 
2 g$, resulting in 6 and 8 particle final states, which have not been constrained by LEP analyses
other than the model independent ones.}
\end{itemize}
Additionally, we can split the scenarios further into two cases: ``large-mixing" and ``small-mixing". We define these as follows: the small mixing case arises when the light CP-even singlet has a sufficiently small Higgs component that s-strahlung limits do not constrain it, even with conventional decays. In the large-mixing case, the singlet has sufficient Higgs component that non-standard decays are needed to evade limits.

With large mixing case, there are three real scalars that we are concerned with: $h$, $s$,
and $a$. These are defined to be the field which couples to the Z in Higgs-strahlung, the
field which mixes with the Higgs, and the field which does not mix with the Higgs,
respectively. In many cases $s$ can be thought of as the scalar and $a$ as the
pseudoscalar. The mass eigenstates are $\tilde h$, $\tilde s$ and $\tilde a$, which are
the fields mostly made of the untilded field of the same character, and usually the
heaviest, intermediate, and lightest mass eigenstates, respectively.  As a note of
caution, we will often refer to the $\fh, \fs, \fa$ still as the Higgs, CP-even singlet
and CP-odd singlet and have tried to ensure that the meaning is clear given the context.

First, we will discuss the necessary operators in all scenarios and then we will proceed to study the three basic scenarios, including the existing limits and model building possibilities.

\subsection*{Model Building: Necessary Operators in All Scenarios}
The basic requirements on any viable scenario are threefold, and can be satisfied with various combinations of operators. The requirements [operators satisfying the requirements] are:
\begin{itemize}
\item{$\fa$ must decay [\ensuremath{A_h}, \ensuremath{m_{CP}^2}, \ensuremath{M_Q^{-1}}]: This is not absolutely true, for instance if \fa\ is never produced, but almost always is.}
\item{\fh\ must decay into 2\fs\ or 2\fa\ or \fa\fs\  [\ensuremath{m_D}, \ensuremath{\lambda_h}, \ensuremath{A_s} (+ mixing), \ensuremath{A_h}]:  For the mass range we are interested in, these non-standard decays for \fh\ must exist.} 
\item{\fs\ must decay to 2\fa\ [\ensuremath{\lambda_h} (+ mixing), \ensuremath{A_s}, \ensuremath{A_h} (+ mixing), \ensuremath{m_{CP}^2}]: When \fs\ is light (i.e. $\lsim 80\; \gev$) and in ``large-mixing'', constraints from SM decays are usually quite stringent. Conceivably, \fs\ could decay directly primarily to $\tau$'s if the mixing is small enough, but in general, the cascade decays for $\fs$ are necessary to evade LEP limits, unless it is heavier than \fh.}
\end{itemize}

\subsection{Higgs Decays Through a Single Stage Cascade Into b Quarks}
\label{sec:mixing}
As we have already described, the only means in the
MSSM to push up the mass of the Higgs above LEP limits is to introduce
very heavy top squarks, which then introduce tunings into the theory.
However, as described in section \ref{sec:singlets}, 
if the Higgs mixes with a lighter field \s, the mass of the field \fh\
which is produced strongly through \fh-strahlung from a Z can have its
mass pushed up, simply by mass mixing, without introducing unnaturally
heavy top squarks. However, this mixing 
allows the lighter field to be produced weakly through \fs-strahlung, trading off the heavier \fh\ for stronger limits on the
lighter \fs. 


Now that \fs\ contains a Higgs component, we must consider LEP limits
on it. 
With standard model decays for \fs, the limits are typically
$\sin^2\theta \lsim 0.03$.  The large mixing
necessary to achieve changes in the \fh\ mass requires
that \fs\ decays dominantly through cascades, in this case, one
typically requires $\sin^2\theta \lsim 0.2$ if the cascades end in b
quarks for lighter \fs\ and $\sin^2\theta \lsim 0.4$ for heavier
\fs.
These, in general, also imply that the dominant decays of
the Higgs are non-standard.

If these decays are $\fh \to 4 \tau$ or $\fh \to 4 g$, then the Higgs
can be considerably lighter than 115 GeV, and it is essentially
unnecessary to push the Higgs mass up. These scenarios will be discussed in section
\ref{sec:nob}.  


We show the allowed regions (dark and light shaded) with \fh\ decays to four b jets in figures
\ref{fig:mixplot}a and \ref{fig:mixplot}b. In these plots, we assume a value for $m_a>12
\gev$ (where kinematics no longer strongly favor decays to $\tau$'s) where constraints are
strongest.
We show the regions which
are allowed when accounting for constraints on $\fs \to 4b$ decays (light and dark shaded regions), assuming a typical $BR_{a\to 2b} \simeq 0.87$. For comparison, we also plot the region where $m_h>110 \ \gev$, where $h\to 4 b$ is essentially unconstrained (dark shaded regions).
\FIGURE[t]{
 \centerline{a)\epsfxsize=2.5 in \epsfbox{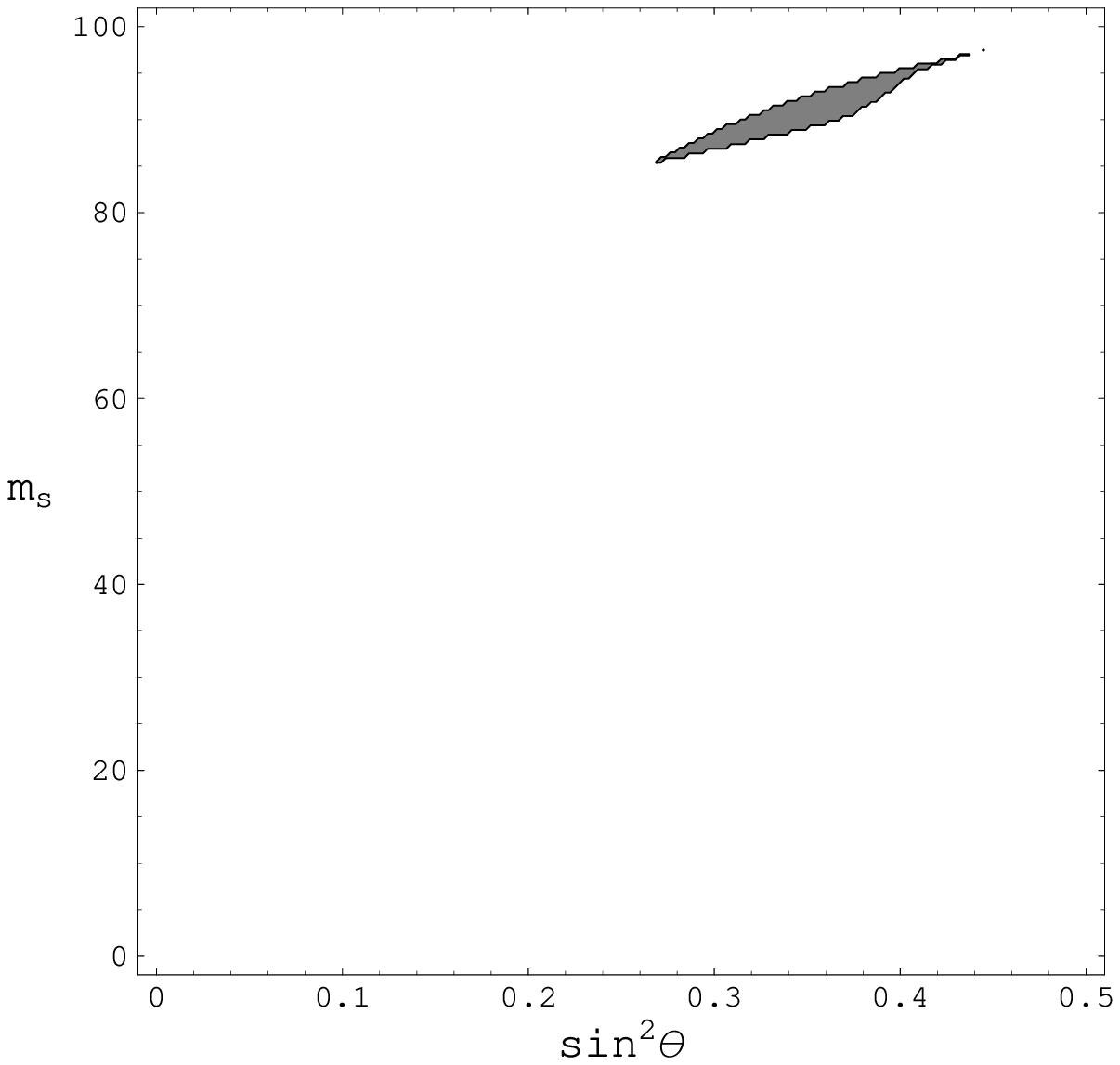}  \hskip 0.5in b) \epsfxsize=2.5 in \epsfbox{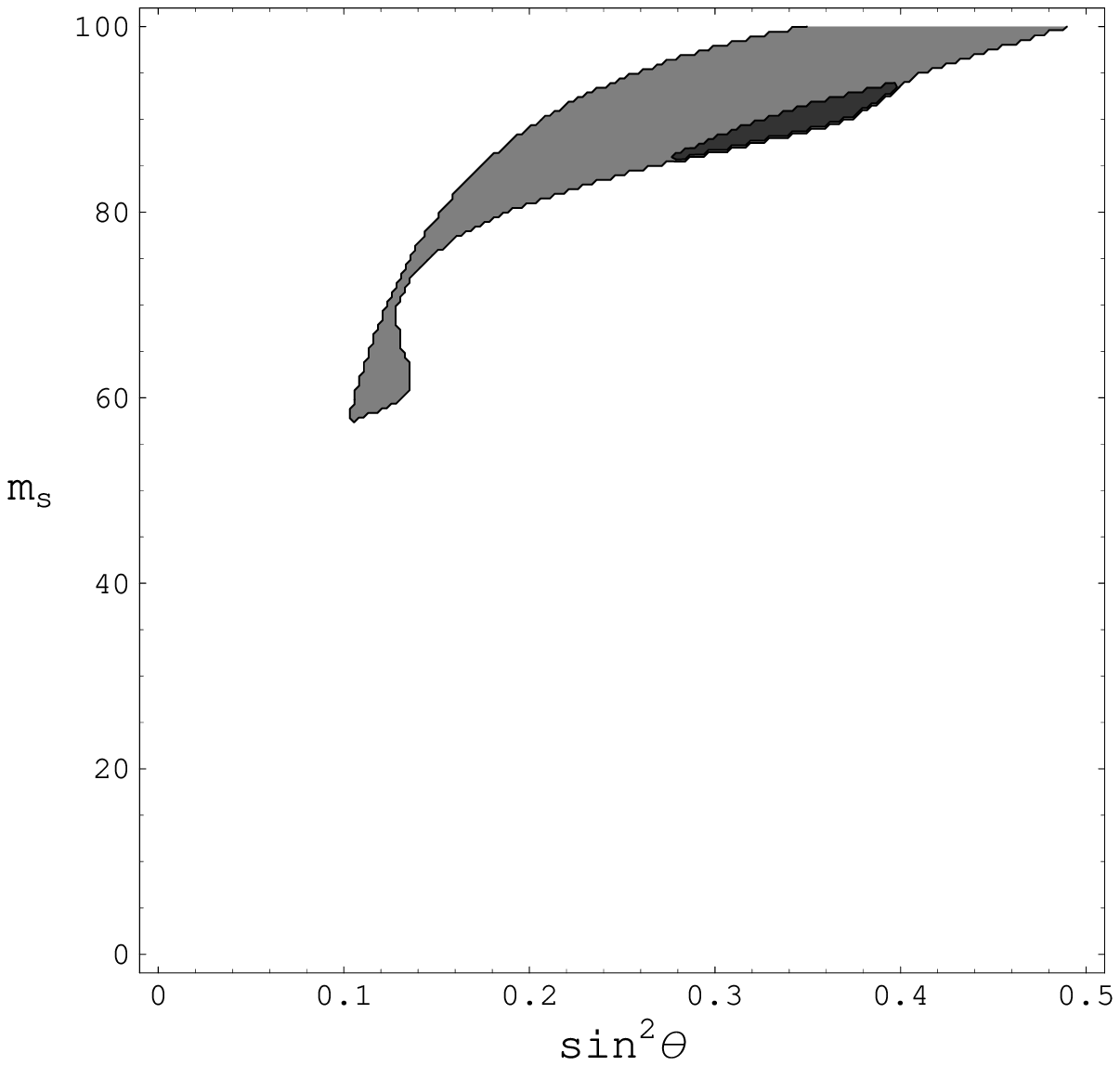}}
\noindent
\caption{The allowed regions in $\sin^2\theta$-$m_\fs$ space, where $m_\fs$ is the mass of a light singlet mixing with the Higgs an amount $\sin\theta$. The plots are shown with $\sqrt{m_{\tilde t_1}m_{\tilde t_2}}=270\ \gev$ (a) and  $\sqrt{m_{\tilde t_1}m_{\tilde t_2}}=300\ \gev$ (b). The shaded regions (combined dark and light) are allowed from the combined LEP limits, while the dark shaded regions additionally have $m_{\tilde h}>110 \ \gev$.
}\label{fig:mixplot}
}

These plots have assumed a 100\% branching ratio for $h \to 2a$, so the only applicable limits are the $h\to 4b$ exclusions given in the LEP-wide analysis \cite{Sopczak:2005su,lepnote}.  However, it is an important question what the LEP constraints are on a scenario with both $h \to 2b$ and $h\to 4b$ rates.  In a recent paper on these decays in the NMSSM \cite{Dermisek:2005gg}, the authors only required that the $h\to 2b\, (4b)$ rate be consistent with the LEP $h\to 2b\, (4b)$ exclusions, thus they assumed that these analyses are independent of each other. However, upon closer inspection of the details of the analyses, it appears such an approach may be too generous. 

In the higher mass ($m_\fh>90 \gev$) range, the combined LEP analyses are dominated by OPAL  \cite{Abbiendi:2004ww} and DELPHI \cite{Abdallah:2004wy}. A close reading of these papers suggests that the $2b$ and $4b$ decays of the Higgs are generally reconstructed together. For instance, in studying $e^+e^-\to Zh$, in the $Z \to \nu \nu$ analysis by OPAL, the neural nets trained to capture to the $4b$ decays of the Higgs are also reasonably efficient in capturing the $2b$ decays. In the all jet ($Z\to jj$) analysis, OPAL uses the same analysis procedure for both $2b$ and $4b$, forcing the six jet event into a four jet topology via the DURHAM algorithm, hence this analysis efficiently reconstructs both types of decays.  As for DELPHI, they clearly state that the analyses are not independent of each other \cite{newdelphi}, and both decays are reconstructed via the same analysis procedure.

Beyond efficiencies, there are still differences between $2b$ and $4b$ events.  
For example, the distribution of discriminating variables will be broader in $4b$ decays (e.g. the reconstructed Higgs mass), making it 
difficult to know how to constrain the scenario when there are significant levels of both $2b$ and $4b$
decays.  If $2b$ and $4b$ signals were indistinguishable, the
correct limit would be $\frac{\xi^2_{4b}}{\xi^2_{4b,bd}}
+\frac{\xi^2_{2b}}{\xi^2_{2b,bd}}< 1$,  where $\xi_{bd}^2$
is the experimental bound of the individual analysis. As an attempt to combine these
limits, when there are $O(1)$ rates for $2b$ and $4b$ decays, in addition to
applying the individual limits, we will also require that
\be
\xi^2_{2b+4b} \equiv \frac{\xi^2_{4b}}{\xi^2_{4b,bd}} +\frac{\xi^2_{2b}}{\xi^2_{2b,bd}}< \sqrt{2}
\label{eq:xieff}
\ee
This additional requirement, on the effective $\xi_{2b+4b}^2$, accounts for the redundancy of the
analyses, and which should forbid situations where the naive combination
of analyses is significantly excluded.  Thus, we find it to be a good compromise between the assumptions of complete independence/interdependence ($\xi^2_{2b+4b} \leq 2,\ 1$ respectively) of the separate analyses.  Note that one can also interpret this as only requiring a 99.5\% CL limit, 
if the two decays give indistinguishable events.   More rigorously, a combined analysis should be done to find the proper constraints.   

For our purposes here, where we take the branching ratio $h \to 2a$ to be one, the results are relatively simple. Even with relatively light stops (270 GeV), the small suppressions of the Higgs couplings due to mixing are sufficient to allow such a Higgs to have been undetected. However, at this mass, this is achieved by strongly mixing and pushing the physical Higgs above 110 \gev (out of the LEP constraints). At $m_{\tilde t}=300\ \gev$, a region opens up below 110 GeV, but the overall allowed space is still quite narrow. Moving to higher Higgs mass (roughly 325 GeV stops) the model independent parameter space opens up significantly (about twice as large as at 300 GeV). However, even with 300 GeV stops, the tuning of $m_Z$ is already expected to be O(15\%), which begins to reintroduce fine tuning from another direction.

Note that in all of these cases we are considering \fs\ to be lighter than \fh. The reason
for this is simple: although mixing the Higgs with a heavier singlet can suppress
couplings of the \fh, if it is heavier than the Higgs, the effect is to push the mass of
\fh\ down, aggravating naturalness issues. Thus, we should view a light \fs\ as a natural
consequence of this model with $m_\fh \sim 110\ \gev$.
Such a scenario is somewhat distinct from previously discussed scenarios \cite{Ellwanger:2005uu,Dermisek:2005gg} because of the presence of the light \fs\ which can be easily produced at rates comparable to that of the SM (in general, roughly a factor of 10 smaller). 

Let us further note that we have been discussing the model independent tuning. Whether or not one can achieve these mixings in a given model without, e.g., tuning to prevent a tachyonic \fs\ is a separate, often more stringent constraint.

As a final comment, we note that these limits are based on the best available limits of $h\to 2a\to 4b$, which are still preliminary. Final limits may further constrain this scenario.

\subsubsection*{Model Building: Single Stage Cascades into b Quarks}
It is straightforward to construct models in which the Higgs is mixed
strongly with a lighter scalar. Because of this mixing, if the Higgs
decays $\fh \to 2 a$, then generally $\fs$ decays to $2a$ as well. 
Here we will discuss the model building and tunings associated with large
mixings, and situations where $\fh\to 2a \to 4b$. 
Situations with $\fh\to 2a \to 4g, 4\tau$, where Higgs mass may be below 110 \gev, will be deferred to subsequent discussions.

There are three operators which can induce significant mixing with the Higgs: \ensuremath{m_D}, \ensuremath{\lambda_h}, and \ensuremath{A_h}. \ensuremath{A_h} principally mixes $s$ with $H^0$ rather than $h$ because $\cos(\alpha+\beta)$ is small, so we focus on the other two.

\ensuremath{m_D} is unique in that while it induces mixing, it also adds a diagonal mass for $s$, so that a tachyon never appears. As a consequence, with this operator, it is simple to get large mixing without having to tune masses to a high degree. In figures \ref{fig:massesformD} we see that one can easily achieve large Higgs masses and large mixings over broad ranges of the parameter space.

\FIGURE[t]{
 \centerline{a)\epsfxsize=3 in \epsfbox{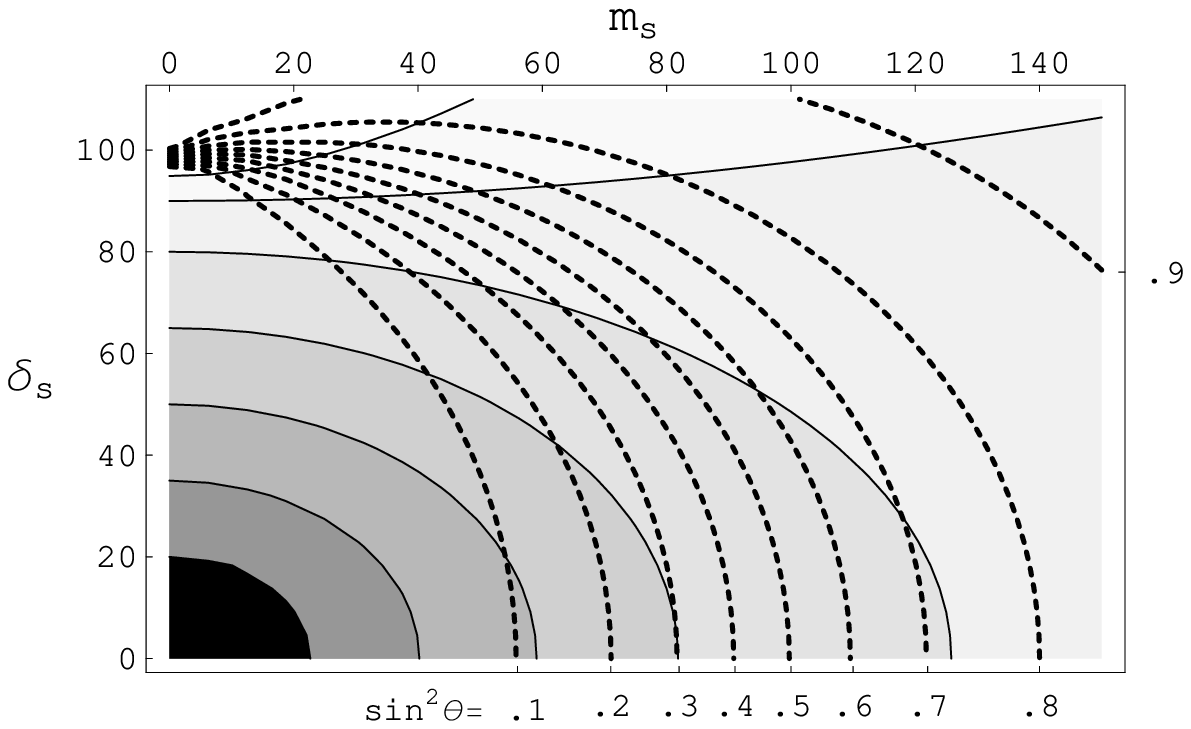}  \hskip 0.2in b) \epsfxsize=3 in \epsfbox{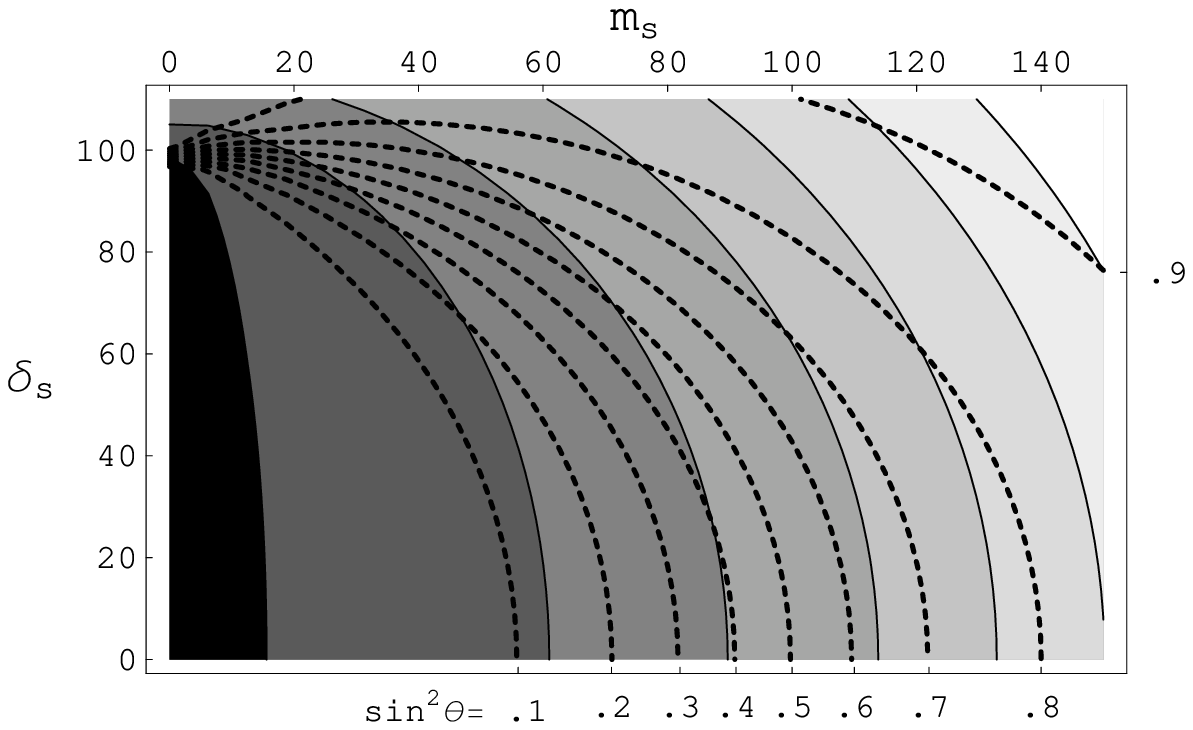}}\\
 \centerline{c)\epsfxsize=3 in \epsfbox{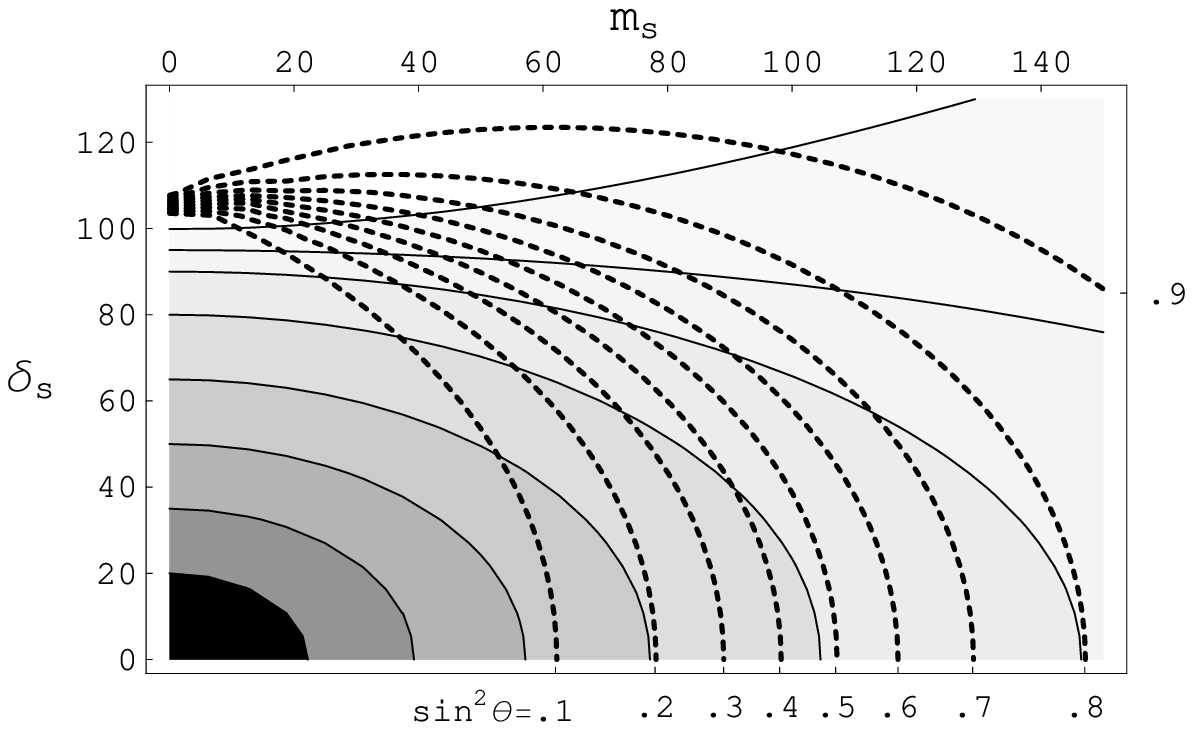}  \hskip 0.2in d) \epsfxsize=3 in \epsfbox{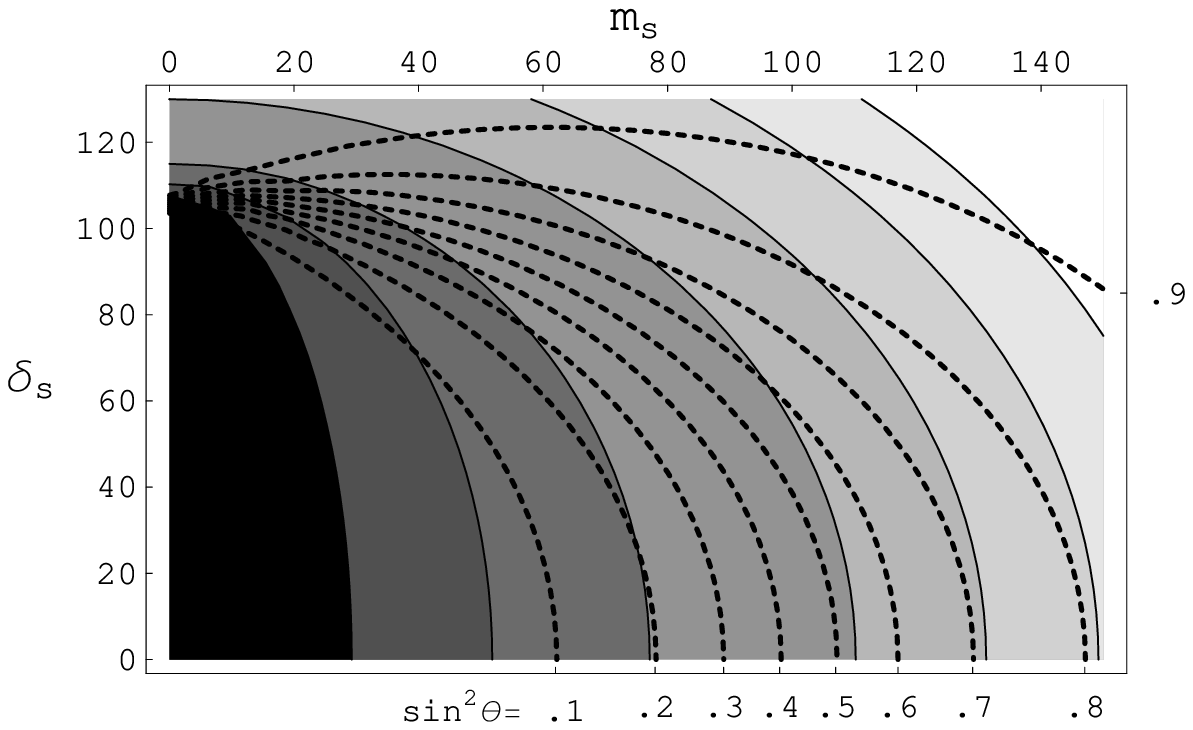}}
\noindent
\caption{In these plots, the contours are of ranges of $\fs$ mass (a,c) or $\fh$ mass (b,d). The x-axis specifies the value of $m_{s}$ arising from the \ensuremath{m_D} operator. The y-axis is all other contributions to the mass squared term for $s$. The values of $\sqrt{m_{\tilde t_1} m_{\tilde t_2}}$ are 250 GeV for a,b and 350 GeV for c,d. The dashed contours are lines of constant $\sin^2\theta$ as labeled. The contours (darkest to lightest) are for masses less than a) 20, 35, 50, 65, 80, 90, 95; b) 100, 105, 115, 130, 145, 160, 175; c) 20, 35, 50, 65, 80, 90, 95, 100; d) 108, 110, 115,130, 145, 160, 175.
}\label{fig:massesformD}
}

\ensuremath{\lambda_h} is somewhat more challenging, because we require a sufficiently large $\mu$-term from experiment. From searches for the chargino, we typically require $\mu > 100 \ \gev$ \cite{Eidelman:2004wy}. In figure \ref{fig:massesforlambdah} we consider this scenario for $\lambda=0.25$. Here we see that it is very challenging to have sufficiently large $\mu$ while keeping the light state dominantly $s$ (i.e., $\sin^2\theta <0.5$). One can tune this scenario to achieve this, but generally, it is most natural to have $\sin^2\theta \sim 1$ and the $s$ decoupled, and $m_h > 110\ \gev$ (as is required with no singlet component). Naturalness here is not considerably improved from the MSSM, unfortunately. 

\FIGURE{
 \centerline{a)\epsfxsize=2.8 in \epsfbox{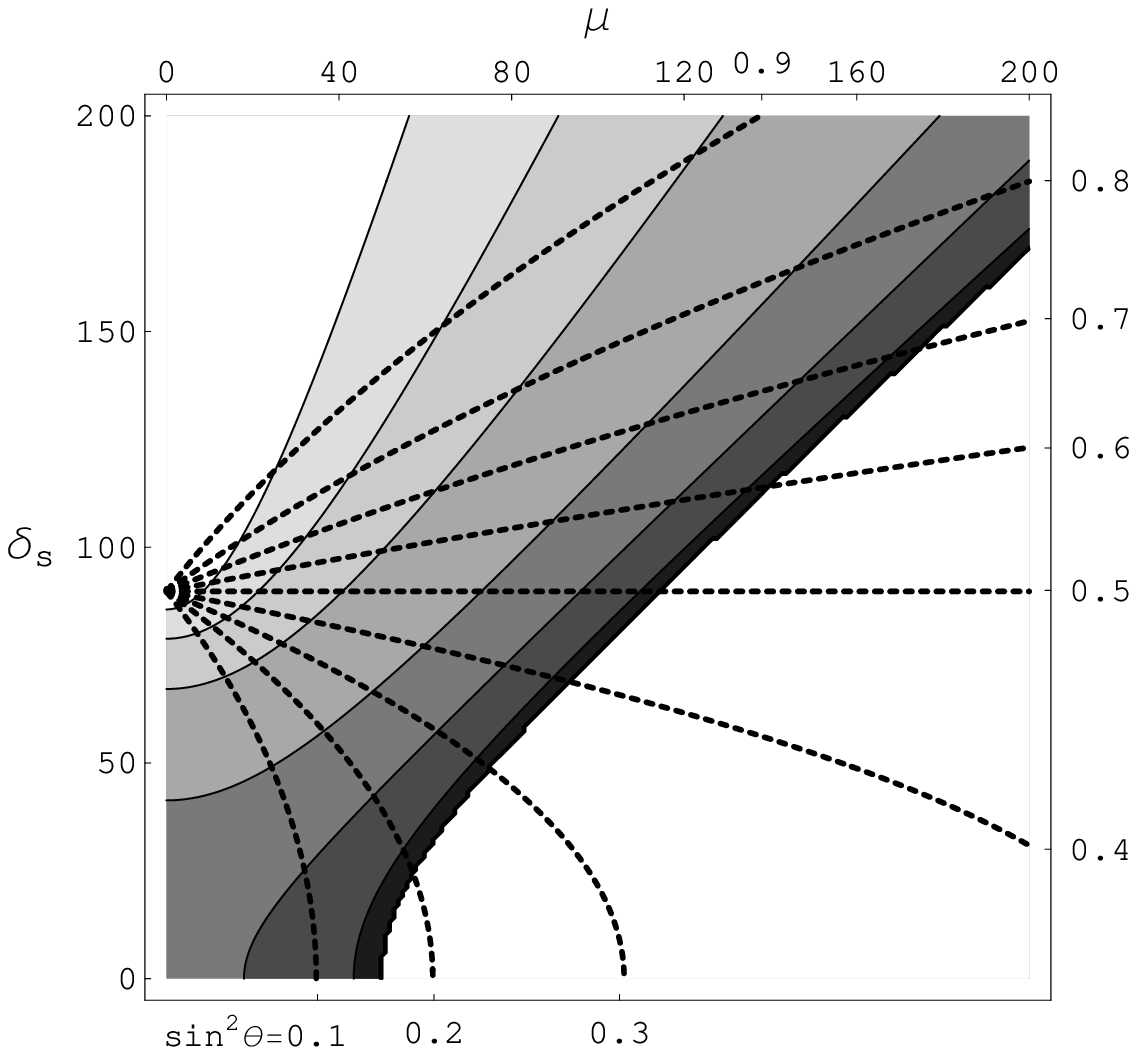}  \hskip 0.2in b) \epsfxsize=2.8 in \epsfbox{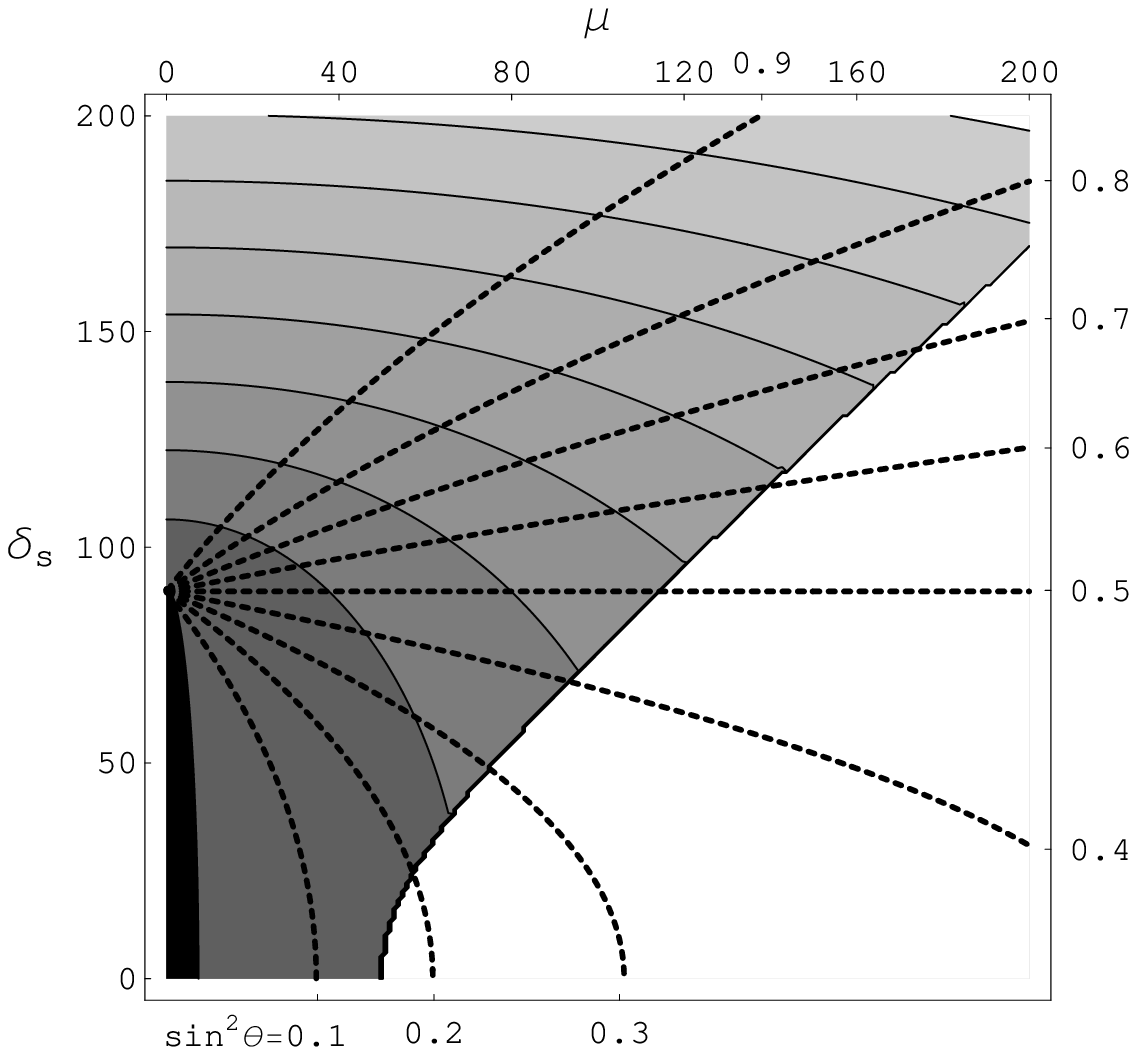}}\\
 \centerline{c)\epsfxsize=2.8 in \epsfbox{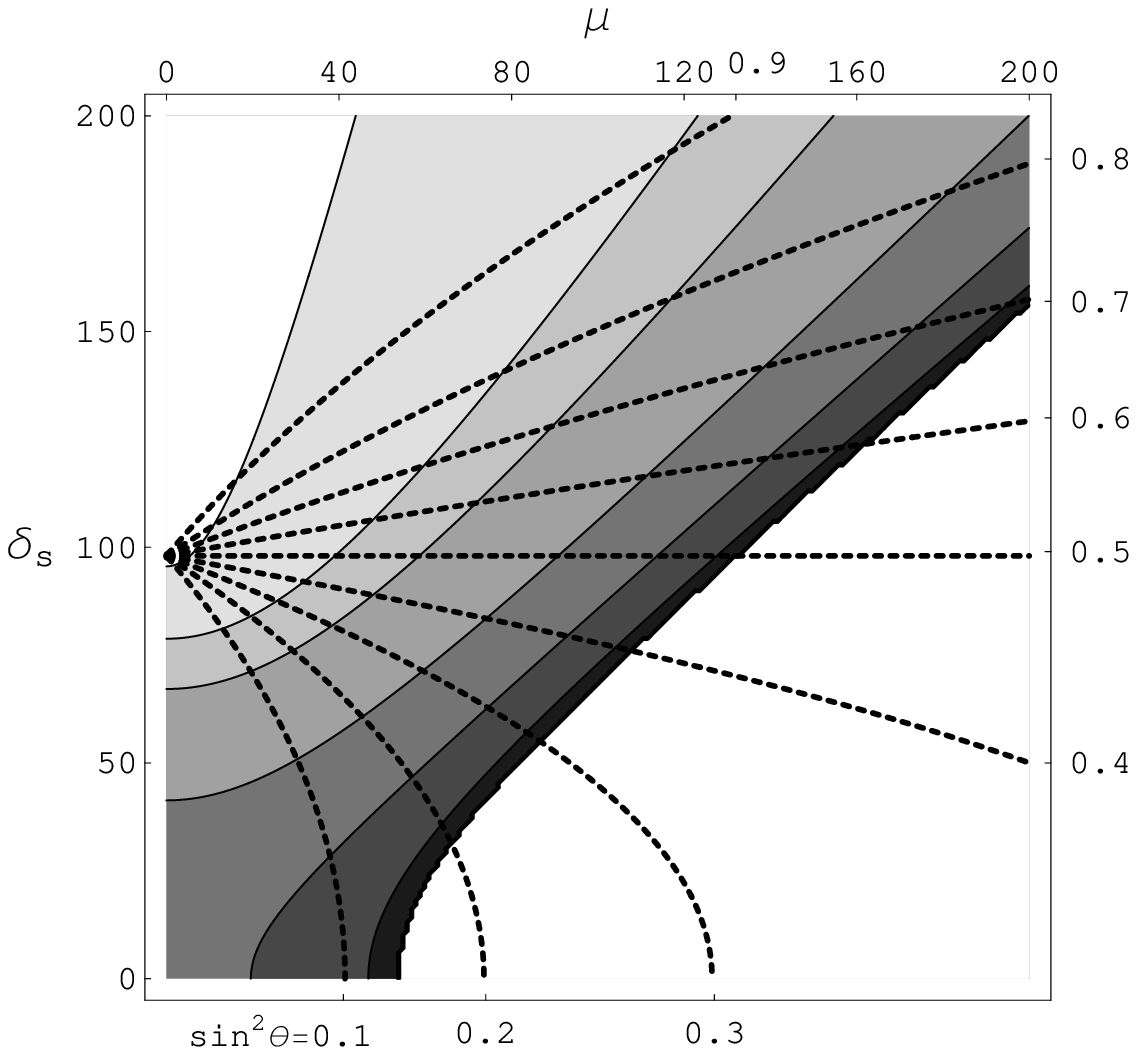}  \hskip 0.2in d) \epsfxsize=2.8 in \epsfbox{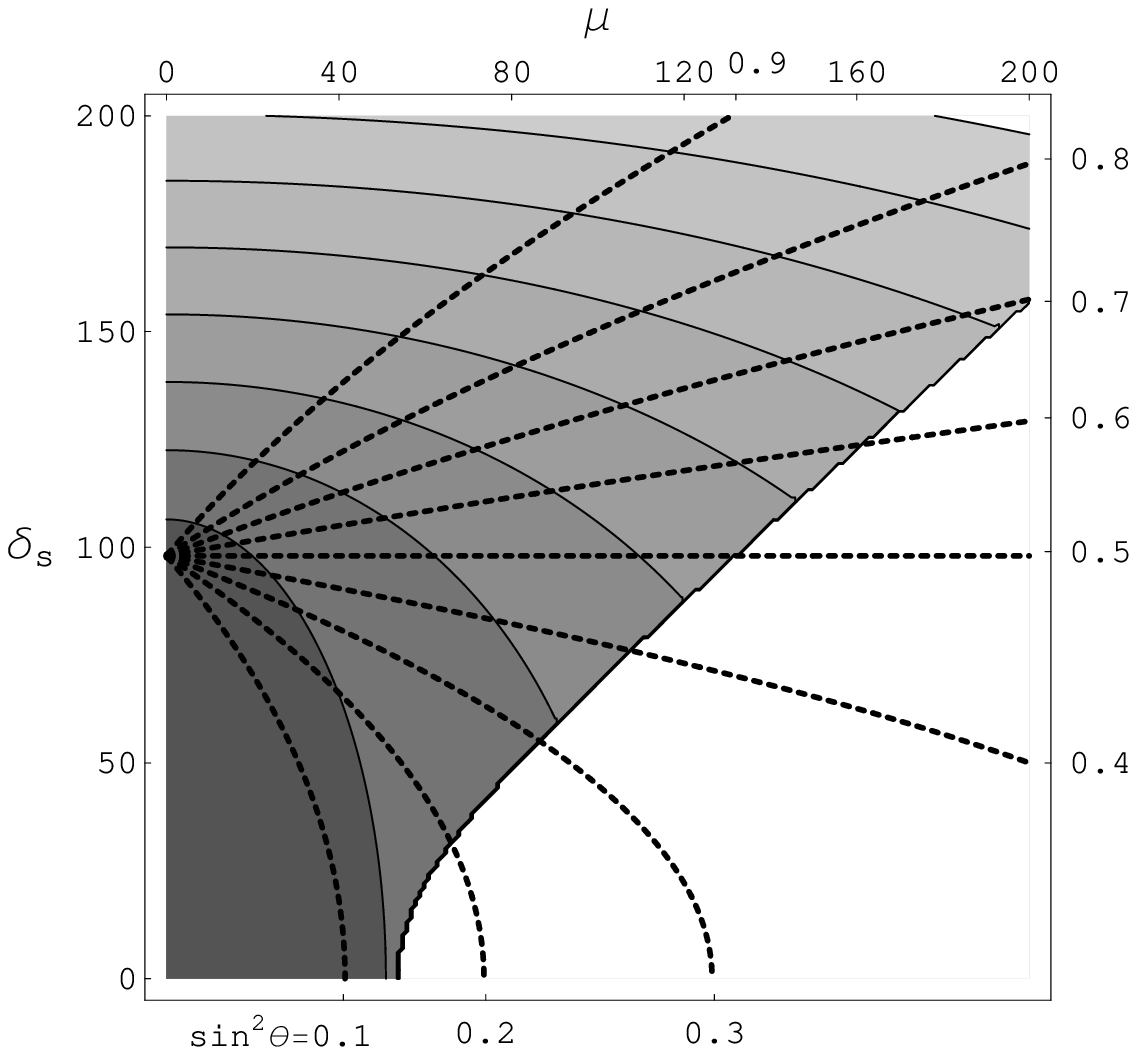}}
\noindent
\caption{In these plots, the contours are of ranges of $\fs$ mass (a,c) or $\fh$ mass (b,d). Here the \ensuremath{\lambda_h} operator is taken to be $\lambda = 0.25$. The x-axis is the value of the effective $\mu$ parameter (once $s$ has acquired a vev). The y-axis are other soft contributions to the mass squared of $s$, in addition to the $\lambda^2 v^2$ term arising from \ensuremath{\lambda_h}. The values of $\sqrt{m_{\tilde t_1} m_{\tilde t_2}}$ are 250 GeV for a,b and 350 GeV for c,d. The dashed contours are lines of constant $\sin^2\theta$ as labeled. The contours (darkest to lightest) are for masses less than a) 20, 40, 60, 80, 90, 96; b) 100, 115, 130, 145, 160, 175, 190, 205, 220; c) 20, 40, 60, 80, 90, 105; d) 105, 115, 130, 145, 160, 175, 190, 205, 220. The white region in the lower right corner is where $\fs$ is tachyonic.
}\label{fig:massesforlambdah}
}

Since \ensuremath{\lambda_h} is in general more tuned, we list a benchmark with
\ensuremath{m_D} below (masses in GeV and in the decoupling limit which we define as $m_A
\gg m_Z$ and large $\tan \beta \gsim 5$) 
\begin{equation}
\begin{array}{|c|c|c|c|c||c|c|c|c|c|c|c|}
\hline
m_{\tilde t} & \ensuremath{\delta_s^2}  & \ensuremath{\delta_a^2} & \ensuremath{m_D} & \ensuremath{A_s} &\sin^2 \theta & m_{\tilde{h}} & m_{\tilde{s}} &m_{\tilde{a}} & B(\fh \to 2\fa)& B(\fs \to 2\fa) &\rm tuning\\
\hline
325 & 45^2 & 57^2 & 43  & -11 & 0.1 & 109 & 73.8 & 32.6 & 0.86 & .99 &3\% \\ 
\hline
\end{array} 
\end{equation}
Here ``tuning'' is defined as the tuning in \ensuremath{m_D} necessary to satisfy the
necessary requirements. The maximum value of \ensuremath{m_D} arises from the limit on
$\tilde s \to 2a$ decays and the minimum value arises from the lower limit on $m_\fh$. The tuning of a parameter $x$ which can lie between two values $a$ and $b$, with $b>a$ will be defined to be the smaller of 100\% or $(b-a)/(b+a)$. Qualitatively, this is the average fractional change in a parameter which is allowed consistent with phenomenology and constraints. In
future benchmarks, it will generally be defined as the largest tuning in
\ensuremath{\delta_s^2} or \ensuremath{\delta_a^2} to achieve the proper spectrum of
$\fs,\fa$ masses.  This tuning is usually related to the lighter of the two states. 
In these scenarios, constraints on $\xi_{2b+4b}^2$ (see eq. \ref{eq:xieff})
are very stringent, so this region may not be allowed by the full analysis.   

\subsection{Higgs Decays Through a Single Stage Cascade (without b's)}
\label{sec:nob}
Single stage cascade decays of the Higgs are familiar within supersymmetric theories.
Within the CP violating MSSM such decays can easily occur
\cite{Pilaftsis:1999qt,Carena:2000ks,Carena:2002bb}. Within the NMSSM, the decays $h\to 2a
\to 4\gamma$ are well studied \cite{Dobrescu:2000jt,Dobrescu:2000yn}, while it has been
argued, prior to the combined LEP analyses, that $h \to 2 a \to 4 b$ allowed considerable
reduction in the fine tuning of $m_Z$ within the NMSSM \cite{Dermisek:2005ar}. More
recently, it has been argued that the combined LEP limits favor a $h\to 2a \to 4b$
decaying Higgs at approximately $m_h \sim 105\ \gev$ \cite{Dermisek:2005gg}, which we argued in section \ref{sec:mixing}, 
had applied constraints that may have been too generous. $h \to 2 a \to 4 \tau$ has also been considered 
\cite{Dobrescu:2000yn,Ellwanger:2005uu}. Here we study
these, and an additional possibility, namely, $ \fh\to 2 \fa \to 4 g$.

\noindent {$\mathbf{ \fh\to 2 \fs, 2\fa \to 4 \tau}$:}
Phenomenologically, it makes little difference whether the intermediate state here is $2\fs$ or $2\fa$, except if it is $2\fs$, the $\fa$ can be completely decoupled, while if it is $2\fa$, the $\fs$ could mix considerably with the Higgs, giving large corrections to the Higgs couplings. In general, models of this sort are highly tuned, because the mass of the intermediate scalar must be $\lsim 10 \,\gev$ in order for decays to $\tau$'s to dominate over decays to b quarks. As a consequence, they generally do not lead to a significantly less tuned theory than the MSSM, although they offer interesting and distinct phenomenology.\footnote{In our scenarios \fh\ decays into 2\fa\ are correlated with the size of the \fa\ mass which leads to tuning for \fa\ to be lighter than two b's.  If the decays are mediated by derivative couplings, e.g. a Peccei-Quinn axion, then this light \fa\ can be reasonably natural (see for e.g. \cite{Hall:2004qd}). \label{foot:pqaxion}} Generically, one expects a scalar field with mass $\lsim 10 \,\gev$, as well as accompanying decays $\fh\to 2\fs, 2\fa \to 2 \tau 2 c$ and $\fh\to 2\fs,2\fa \to 4 c$.  There may also be some interesting B physics in this mass range \cite{Hiller:2004ii}.

The only existing analyses on this scenario \cite{Abbiendi:2002qp,Sopczak:2005su,lepnote} cut off at $m_\fh >86 \,\gev$, for theoretical reasons related to the MSSM which do not exist in theories with additional singlets. The extension of this analysis is important to understanding the true limits on such a scenario.

\noindent{ $\mathbf{\fh\to 2\fa \to 4 g}$:}
One of the most interesting possible signatures results in the 4 glue
final state. Such a final state is difficult to see at LEP and has not been constrained by any existing analysis, save the model independent analysis \cite{Abbiendi:2002qp}, allowing such a Higgs, in principle, as light as $82 \
\gev$. The extent to which we believe other analyses (e.g., the two-jet flavor independent decays) can be interpreted to constrain this scenario will be discussed in the benchmark summary. 

Models with this final state can be easily constructed with
essentially no fine-tuning, and squarks as light as the direct
experimental limit. Although a search for four jets at a hadron collider seems impossible, the same processes which generate the four jet final state can also generate $\fh \to 2 \fa\ \to 2g 2 \gamma$, which should occur with a rate down by roughly $\alpha^2_{EM}/\alpha_s^2\sim 10^{-2}$, see $M_Q^{-1}$ subsection of section \ref{sec:singlets}. This rate can be increased if couplings of the $a$ to Higgsinos exist as well. A search for the two photons at the LHC should be feasible \cite{Dobrescu:2000jt}, and there may be background reduction by requiring two jets which reconstruct to the invariant mass of the photon pair (however, this is unlikely at a hadron collider like the LHC). 


%

\subsection*{Model Building: $\mathbf{\fh \rightarrow 2 \fa \rightarrow 4 \tau, 4g}$ via \ensuremath{\lambda_h}, \ensuremath{M_Q^{-1}}}
\ensuremath{\lambda_h} allows decays to both \fs\ and \fa. However, \ensuremath{\lambda_h} generally mixes the CP even $s$ with $h$, giving strong constraints from \fs-strahlung. Consequently, with \ensuremath{\lambda_h}, the most natural models are those with $\fh \to 2 \fa \to X$. 
In order to evade constraints on \fs-strahlung, it is most straightforward to simply raise the mass of $\fs$ through the addition of $\ensuremath{\delta_s^2}$. In doing so, however, one generally realizes the large mixing scenario described in section \ref{sec:mixing}, requiring that $m_\fs > 2 m_\fa$ in order to evade constraints. One can raise the mass of $\fs$ much larger than the other relevant scales, as well. In fact, this is quite reasonable, for instance if $S$ appears as the N=2 completion of a 5-D $U(1)$ gauge multiplet. 5-D gauge invariance protects the mass of $a$, but not $s$ against tree level SUSY breaking terms \cite{Arkani-Hamed:2003wu,Kaplan:2003aj}. Thus, for our purposes in this section, we will decouple $s$ from the theory.

For $\fh \to 2 \fa$ to sufficiently dominate over $\fh \to 2b$, a suitably large \ensuremath{\lambda_h} is necessary, and by the discussion in section \ref{sec:singlets}, we see that the requirement is $\lambda > 0.2$.
At tree level, this results in mass squared of  $(35\ \gev)^2$ for $\fa$. 

Let us now consider the subsequent decays of $\fa$. If we include the a nonzero \ensuremath{M_Q^{-1}}, the possibility of $\fa \to 2 g$ opens up. However, 
the presence of \ensuremath{\lambda_h} requires the presence of \ensuremath{A_h}, which allows $\fa\to 2 b$ as well. This operator need not be large, being generated at the loop level. At this size
\be
\ensuremath{A_h} \sim \frac{\lambda g^2}{16 \pi^2} m_{susy} \sim .2 \,\gev
\ee
it would be expected to yield a mixing of $a$ and $A^0$ at the $10^{-3}$ level or smaller, which is easily small enough for the decays to glue to dominate.  In comparison, in the NMSSM it is difficult to have \ensuremath{M_Q^{-1}} induced decays dominate because the cross term in $|\kappa S^2 + \lambda H_u H_d|^2$ acts like \ensuremath{A_h} with an $S$ vev and thus fermion decays tend to dominate in this scenario.

The benchmark point then has
(taking the decoupling limit)
\begin{equation}
\begin{array}{|c|c|c|c|c|c||c|c|c|c|c|c|c|}
\hline
m_{\tilde t} & \ensuremath{\delta_s^2}  & \ensuremath{\delta_a^2} & \ensuremath{\lambda_h}
& \ensuremath{A_h}&M_Q^{-1}&\sin^2 \theta & m_{\tilde{h}} & m_{\tilde{s}} &m_{\tilde{a}} &
B(\fh \to 2\fa) & \rm tuning\\
\hline
200 & 500^2  &-35^2 & 0.3 & 0.3& 300^{-1}&10^{-3} & 92.1 & 503 & 38.7 & .92 & 60\% \\
\hline
& &(-51^2) & & &(0) & &  & &(11.1) & (0.96) & (3\%)\\
\hline
\end{array} 
\end{equation}
Here we have taken 
$\mu=100\ \gev$, there is no reduction in the effective $\mu$-term since
$\delta_s$ is large.
The inclusion of a nonzero \ensuremath{M_Q^{-1}} as described in subsection \ensuremath{M_Q^{-1}} of section \ref{sec:singlets} generates the decays to two gluons.  Due to the kinematics (heavy \fh, \fa), there are no constraints on these 4 jet decays.

If \ensuremath{M_Q^{-1}=0}, one expects decays to standard model fermions to dominate. 
For decays of taus to dominate requires $m_\fa$ to be less than the $b\bar{b}$ threshold.
 This requires us to tune \ensuremath{\delta_a^2} to cancel the tree-level mass of $m_a^2
\sim (40 \gev)^2$. For instance, the above point with \ensuremath{\delta_a^2} of $-(51
\gev)^2$ achieves the proper spectrum with a 3\% tuning.  The
necessary changes to the parameters and the corresponding changes in the spectrum are
shown in parentheses in the table in the second line, whereas things that are common to
the first are not repeated.

\subsection*{Model Building: $\mathbf{\fh,\fs\ \to 2\fa\ \to 4g, 4\tau}$ via \ensuremath{m_D}, \ensuremath{A_s}, \ensuremath{M_Q^{-1}}}
Cascade decays of the CP-even states into $2\fa$ can easily dominate with the inclusion of \ensuremath{m_D} and \ensuremath{A_s}.  
The operator \ensuremath{A_s} induces a large trilinear $s a^2$ while \ensuremath{m_D} induces $O(1)$ mixing between $s$ and $h$. Thus 
if the $2\tilde{a}$ channel is kinematically accessible, $\tilde{h},\tilde{s}$ will dominantly decay into it.  In addition, 
there is no mixing in 
the CP-odd sector, thus \ensuremath{M_Q^{-1}} provides the only decay channel for $\tilde{a}$; hence the final state of the cascade decay is 
dominantly $4 g$ and, at a subdominant
but potentially interesting rate, $2 g\, 2\gamma$ and $4\gamma$.  Notice that in this scenario the vev of $s$ has no relation to the $\mu$ parameter and the minima are only local minima due to the \ensuremath{A_s} operator.   

As a benchmark point, we take (in the decoupling limit)      
\begin{equation}
\begin{array}{|c|c|c|c|c||c|c|c|c|c|c|c|}
\hline
m_{\tilde t} &\ensuremath{\delta_s^2} & \ensuremath{\delta_a^2} & \ensuremath{m_D} &
\ensuremath{A_s}  &\sin^2 \theta & m_{\tilde{h}} & m_{\tilde{s}} &m_{\tilde{a}} & B(\fh \to 2\fa)& B(\fs \to 2\fa)&
\rm tuning\\
\hline
175 & 80^2 & 0\ & 30 & 8  & .22 & 94.9  & 76.2  & 28.3\ & .92\ & .99
& 100\% \\
\hline
 &  & (-27^2) &  &   &  &   &   & (8.37) &  (.93) & 
& (10\%) \\
\hline
\end{array} 
\end{equation}
Some comments on this benchmark are in order.  Looking at the first line, it is a large mixing point, where $\fs,\fh$ are similar in mass
and the main constraints are on the $\fh$ decays. As mentioned above, with no \ensuremath{A_h} operator, 
there is no mixing in the CP-odd sector so  \fa\ can only decay into gauge bosons through \ensuremath{M_Q^{-1}}.  The constraint from the
dominant $4g$ final state only requires $m_{\fa} \gsim 5\ \gev$ \cite{Abbiendi:2002in}, so these scenarios are viable and quite natural.  

As a variation on the phenomenology, $\tau$ final states can also be considered.  
For the $4\tau$ state to compete, one could include a small \ensuremath{A_h} operator and
tune the \fa\ mass below $b\bar{b}$ threshold, at the expense of increased tuning.  As
before the necessary changes in parameters are given in parentheses in the second line.
The constraints on cascades into $\tau$ now apply, which generally require that
$m_{\fa} \gsim 5\ \gev$ if $B(\fa\ \to 2\tau) \sim 1$.
However, these points are on the edge of a constraint contour, for instance when $m_{\fs}$
is between 45-76 GeV, one instead requires $m_{\fa} \gsim 7\ \gev$, severely limiting the
allowed parameter space \cite{Abbiendi:2002in,Sopczak:2005su,lepnote}.  
In summary, these benchmark points suggest that cascades into $\tau$'s 
seem to be disfavored by both tuning considerations and available parameter space, but still remain an interesting
phenomenological possibility.

\subsection{Higgs Decays Through a Two Stage Cascade}
A very interesting phenomenological possibility is that the Higgs cascade decay proceeds
through two intermediate scalars. This is a very reasonable possibility in supersymmetric
theories, because the new singlet actually comes with both the scalar and pseudoscalar,
so $\fh\to 2\fs \to 4\fa \to 4X$ is possible or with CP mixing, $\fh \to \fs \fa \to 3
\fa \to 3X$ is possible. However, when considering theories with \ensuremath{\lambda_h},
it is essentially impossible to achieve two stage cascades as the dominant
decay mode, see Fig. \ref{fig:massesforlambdah}. Furthermore, with \ensuremath{\lambda_h},
$\fs$ is lighter than $\fa$ due to mixing, 
making it difficult to arrange a scenario
where something of the form $\fh \to 2\fs \to 4\fa \to 4X$ can occur.  

However, in the broader context we are considering here, such decays can easily arise,
with only moderate spectral tuning. In general, the models rely upon the presence of
D-term supersymmetry breaking through the supersoft operator \ensuremath{m_D}.

There are two forms which we can study, one with CP mixing and one without. With CP
mixing, we can consider $\fh \to \fa\fs \to 3 \fa \to 6\tau, 6b$. Such scenarios can
easily occur with essentially no tuning, and occur when the decay $\fh \to 2\fs$ is
kinematically forbidden. In cases where the six b-quark final state dominates, we also
expect a non-trivial fraction of $4 b\, 2 \tau$ events as well.

Without CP mixing, the cascade generally follows $\fh \to 2\fs \to
4\fa \to 8 g, 8 \tau, 8b$. Scenarios with so many soft final states
have not been studied at LEP, and thus have no constraints beyond the
OPAL model independent constraints $(m_h> 82 \ \gev)$. This allows
very light stops, and the only tuning required is in achieving the proper
spectrum of scalars.

In cases with eight gluon jets, there are typically levels of $6g 2 \gamma$ which are non-negligible. However, the photons are sufficiently soft that the background is very large, and the jets are very soft, well below the cuts to be applied at the LHC. Likewise, the decays to eight b quarks or eight $\tau$'s seem difficult to study at hadron colliders.  These scenarios probably will instead be studied by a future linear collider like the ILC.

\subsection*{Model Building: $\mathbf{\fh\ \to 2\fs\to 4\fa\ \to 8g,8\tau,8b}$ via \ensuremath{m_D}, \ensuremath{A_s}, \ensuremath{M_Q^{-1}}}
With the operators \ensuremath{m_D},\ensuremath{A_s}, and \ensuremath{M_Q^{-1}}, it is possible to find
points that have significant two stage cascades.  With the following benchmark point, the Higgs-like mass eigenstate is
pushed heavy by mixing with the $s$, and via dominance of \ensuremath{m_D} over \ensuremath{A_s}, decays $\fh \to 2 \fs$ are favored over
decays into 2\fa.  The small value of \ensuremath{A_s} is still sufficient to have \fs\ decay into 2\fa 's which completes the two stage cascade.  
\begin{equation}
\begin{array}{|c|c|c|c|c||c|c|c|c|c|c|c|c|}
\hline
m_{\tilde t} & \ensuremath{\delta_s^2} & \ensuremath{\delta_a^2} & \ensuremath{m_D} &
\ensuremath{A_s}  &\sin^2 \theta & m_{\tilde{h}} & m_{\tilde{s}} &m_{\tilde{a}} &B(\fh\ \to 2 \fa)& B(\fh\ \to 2
\fs)& B(\fs\ \to 2\fa)&{\rm tuning}\\
\hline
360 &40^2 & -52^2  & 60 & 8  & .06 & 111 & 39.3 & 16.2\  & .35\  & .50\
  & .99 & 4\% \\
\hline
&  &  (-54^2) & &   &  & & & (7.13) & (0.36) &  (0.49) &  &   (2\%)\\
\hline
\end{array} 
\end{equation}
At this point, the constraints that are independent of the \fa\ decay products are: 1) decay independent analysis on \fs\ 
which requires $m_{\fs} \gtap 5 \gev$ for
the value of $\sin^2 \theta$, 2) SM searches on the \fs\ and \fh\ which are both satisfied with stronger constraints on \fh\ since 
$\xi^2_{SM} = \cos^2 \theta B(\fh \to SM) = .14$.

The final state products of \fa\ determine what additional constraints apply to this scenario.  In the first line of the benchmark, we will consider gluon and $b$ quark final states.  
With \ensuremath{A_h} turned off,  a nonzero \ensuremath{M_Q^{-1}} causes it to decay primarily to gluon jets.  For this final state, there are no further analyses that
constrain this scenario.  However, if the \fa\ does mix with $A^0$ by both turning on a small \ensuremath{A_h} operator and reducing \ensuremath{M_Q^{-1}}, 
decays $b$'s dominate.   In this case, the limits apply to the effective $\xi^2_{2b+4b}$ defined earlier.  We find that these are consistent with the new LEP limits, assuming that the analyses are not sensitive to the $8b$ final state (this sensitivity has not yet been determined by any LEP analysis).   
If \fa\ is lighter than 10 GeV, again listed in the second line with new parameters/changes in parentheses, $\tau$'s can dominate, and is consistent with the 
new LEP analysis of $4\tau$ events from \fs -strahlung and Higgs-strahlung.   
Finally, in the case that \ensuremath{M_Q^{-1}} and mixing decays are comparable, more complicated scenarios are allowed, with many possible final state
topologies.

\subsection*{Model Building: $\mathbf{\fh\to \fa\fs \to 3\fa \to 6b,6\tau}$ via \ensuremath{m_{CP}^2}, \ensuremath{m_D}, \ensuremath{A_s}}
In the presence of \ensuremath{m_{CP}^2}, \ensuremath{m_D} and \ensuremath{A_s}, the coefficient of the $\fh \fs \fa$ coupling is 
\bea
\frac{c v}{\sqrt{2}} = \frac{3 {m^2_{CP}} v \Delta  s^2_{2 \theta} \left(m_{\tilde h}^2-m_{{\tilde{s}}}^2\right)}{8 \sqrt{2} m_{\tilde h}^2 m_{{\tilde{s}}}^2}-\frac{{m^2_{CP}} s_{ 2 \theta } g_Y \left((3 c_{2 \theta}+1) m_{{\tilde{h}}}^2+(1-3 c_{2 \theta }) m_{{\tilde{s}}}^2\right) m_D}{8 m_{{\tilde{h}}}^2 m_{{\tilde{s}}}^2}\\ \nonumber +\frac{{m^2_{CP}}\ {A_s} s_{2 \theta } \left((3 c_{2 \theta} )+5) m_{{\tilde{ h}}}^2+(5-3 c_{2 \theta} ))
   m_{{\tilde{s}}}^2\right)}{2 m_{{\tilde{h}}}^2 m_{{\tilde{s}}}^2}
\eea
Where we have taken the decoupling and large $\tan\beta$ limits, as well as the $m_\fa=0$
limit and at tree-level $\Delta=g_Y^2+g^2$.

We are most interested in scenarios where $2m_\fa <  m_\fs \sim 60\, \gev$. Using the perturbative expression for $\sin \theta_{ah}$ (eq. \ref{eq:amix}), we require $\sin^2 \theta_{ah}<0.015$, which is a typical upper limit for $\fa \to 2 b$. The constraints on $\fa \to 2 \tau$  are roughly $\sin^2 \theta_{ah}<.07$ in the kinematically allowed region. Comparing this limit with eq. \ref{eq:amix}, we have (neglecting terms down by $m_\fs^2/m_\fh^2$)
\be
m^2_{CP} \lsim .12 \ m_\fs^2/(\sin \theta \cos \theta) \ee
for b quarks, and 
\be
m^2_{CP} \lsim .26 \ m_\fs^2/(\sin \theta \cos \theta) \ee
for tau decays. Hence, $m^2_{CP}$ can be $O(m_{\fs}^2)$, and thus not a small parameter.
One can still use the approximate expressions for estimates and intuitive understanding,
but for better than O(1) precision, we must calculate exactly. 

The presence of \ensuremath{A_s} is important, both for $\fh\to \fa\fs$ and $\fs\to 2\fa$ decays. In general, one does not need a large value for \ensuremath{A_s}, with $4\ \gev$ sufficient in order to secure sufficiently large $\fh\to \fa\fs$ and $\fs\to 2\fa$ decays.

Let us consider the benchmark point (mass units are in GeV):      
\begin{equation}
\begin{array}{|c|c|c|c|c|c||c|c|c|c|c|c|c|c|}
\hline
m_{\tilde t} & \ensuremath{\delta_s^2} & \ensuremath{\delta_a^2}& \ensuremath{m_D} &
\ensuremath{A_s}  & \ensuremath{m_{CP}^2} &\sin^2 \theta_{sh} & \sin^2\theta_{ah} & m_{\tilde{h}} &
m_{\tilde{s}} &m_{\tilde{a}} & B_{\fh \to \fa \fs}& B_{\fs \to 2\fa}& \rm tuning\\
\hline
250 & 58^2 & -10^2  & 48 & 6  & 42^2 & 0.10 &.01 & 103 & 67.0 & 18.4 & .70 & .91 & 100 \%
\\
\hline
& & (-20^2) & &   & &  & & & (66.6) & (9.87)&  (0.69) & (0.96) & 18 \% \\
\hline
\end{array} 
\end{equation}
In line one, we consider $b$ quark decays (gluon decays cannot dominate since there is mixing between $a$ and $h$).  This point is consistent with the limits on the effective $\xi^2_{2b+4b}$ from $4b$ and $2b$ decays.  To get to $\tau$ decays, one changes the parameters as listed in line two, which satisfies the constraints on $4\tau$ rates.  Again, these results assume limited sensitivity at LEP to the 6 final state decays ($6b$ and $6\tau$).


\section{Phenomenology and Benchmark Point Summary}
\label{sec:benchsum}
It is remarkable that by simply extending the MSSM to include a singlet superfield, there can be so many modifications to the Higgs phenomenology. Furthermore, many of these modifications do not arise in the NMSSM framework.

To aid in understanding, we have attempted to isolate a few points in parameter space which demonstrate the relevant phenomenology. There are several qualitative issues which are relevant. Let us begin by reviewing our different decay scenarios.
\vskip 0.1in
{\noindent \bf i)
$\mathbf{\fh \to 2\fa \to 4 b}$}: The new LEP limits are noticeably more constraining, and
require a certain degree of suppression in the Higgs production and branching ratio of
$\fa$ to b quarks. This can be achieved with $\sqrt{m_{\tilde t_1}m_{\tilde t_2}} = 300
\gev$ ($O(10-15\%)$ tuning).   In a crude attempt to combine limits on $2b$ and $4b$ decays, 
we defined a cumulative $\xi^2_{2b+4b}$ (in eq. \ref{eq:xieff}) and required it to be less than $\sqrt{2}$.  
This is allowed principally in the ``just so'' region,
where we have essentially saturated the LEP limits, i.e., pushed the parameters so that
LEP was nearly sensitive to this scenario.  Hence, these regions may be impacted by the
finalized LEP limits. Hidden tunings seem to appear when we attempt to use
\ensuremath{\lambda_h}, the operator used in the NMSSM, to generate the decays because of
the large $\mu$ term and the subsequent large mixing term with $\fs$. Such hidden
tunings are smaller in models with the supersoft operator \ensuremath{m_D} together
with \ensuremath{A_s}.

\vskip0.1in
{\noindent \bf ii) $\mathbf{\fh \to 2 \fs/2\fa \to 4 \tau}$}: We have
not demonstrated a model which can achieve these decays above the
kinematical threshold (i.e., where $\fa\to bb$ is not
kinematically forbidden). However, we can achieve these decays by tuning the models
to the 10 percent level (necessary to get the $\fa$ mass into this
region). 
Furthermore, it is troubling that
the best available limits in this region (from OPAL
\cite{Abbiendi:2002in}) seem to stop at 86 GeV, a theoretical prior due to constraints
in two Higgs doublet models which does not apply in cases with
singlets. This motivates a reanalysis of the LEP data without the
theoretical bias that $m_\fh<86 \ \gev$ if $m_\fa \lsim 10 \ \gev$. 

\vskip0.1in
{ \noindent \bf iii)  $\mathbf{\fh \to 2 \fa \to 4 g}$}: It is
remarkably simple for this decay to dominate under the assumption that
\ensuremath{A_h} is loop suppressed or absent. Such models could be very natural with
arbitrarily light stops (subject to direct search limits), and a Higgs
with mass $m_\fh$ as light as 82 GeV, the limit from the OPAL model
independent analysis \cite{Abbiendi:2002qp}. These scenarios can arise
easily with either \ensuremath{m_D} or \ensuremath{\lambda_h}, and generally come with associated
$\fh\to 2 \fa \to 2g\, 2 \gamma$ decays at the $7 \times 10^{-3}$ level.

\vskip0.1in
{\noindent \bf iv)  $\mathbf{\fh \to \fa\fs \to 3 \fa \to 6 b, 6\tau}
$}: Such a decay can dominate Higgs decays via the inclusion of the
\ensuremath{m_D} operator, incorporating the sensitivity of the Higgs to D-term
supersymmetry breaking, but is difficult to engineer with only \ensuremath{\lambda_h}
(i.e., without D-term breaking). This is an important example of
phenomenology which would not occur in the NMSSM, but easily arises
within a general operator analysis. Here there is no analysis to
exclude this scenario, meaning a Higgs as light as 82 GeV is allowed.
In the case that the final state is composed of b-jets, such a
scenario can be quite natural. In the case where the final state is
composed of $\tau$'s, the tuning is significant, again at the few
percent level. Such a scenario would be difficult to detect at the
LHC.  Finally, although this scenario seems CP-violating, it can be consistent 
for $\fa,\fs$ to be both CP-even and thus there are no additional contributions
to CP-violating observables such as edms.      

\vskip0.1in
{\noindent \bf v) $\mathbf{\fh \to 2 \fs \to 4 \fa \to 8 b, 8\tau , 8g}
$}: This scenario, too, can only dominate with D-term breaking, and
not within the narrow NMSSM framework. With D-term breaking, it
appears to be necessary to have roughly $10\%$ tuning in order to
achieve the proper spectrum (with b or g final states - a few percent
if the final state is $\tau$), but again a light Higgs (as light as 82
GeV) is allowed (however, in our benchmark, an additional $4b$ rate pushes up the required Higgs mass). Because the final state particles are so soft, it is
difficult to envision a scenario in which the LHC could detect this
Higgs.

The benchmarks realizing these phenomenologies are summarized in Table \ref{table:benchmark}.  The benchmark points illustrate the importance of the general operator analysis. Some scenarios are only natural with the presence of D-term SUSY breaking. The gluon final states only occur once we additionally consider the effects of new, heavy fermions. Such effects are typically excluded from NMSSM analyses and yet we see they can generate some of the most interesting phenomenology.  Despite this, one should keep in mind that unexplored operators may also generate the same phenomenology, and thus the phenomenology presented should not only be considered in the context of the particular model realizations.
\TABLE[t]{\begin{math}
\nonumber \begin{array}{|c||c|c|c|c|c|c|c|c|c|}
\hline
\rm Case &           1 & 2 &3 & 4 & 5 
\\ \hline \hline
\sqrt{m_{\tilde t_1} m_{\tilde t_2}}(\gev)&325    & 200      & 175    &    360   & 250
\\ \ensuremath{\delta_s^2} (\gev)^2      & 45^2  & 500^2 & 80^2       &	 40^2     & 58^2
\\ \ensuremath{\delta_a^2} (\gev)^2      &  57^2 & -35^2 (-51^2) & 0\ (-27^2) &  -52^2\
(-54^2) &
 -10^2\ (-20^2)
\\ \ensuremath{m_D} (\gev)& 43 & 0 & 30             &	60         &  48
\\ \ensuremath{\lambda_h} & 0 & 0.3 & 0 					&  0		& 0
\\ \ensuremath{A_s} (\gev)& -11 & 0 &	8				&   8 	&  6
\\ \ensuremath{A_h}(\gev)& 0 & 0.3 &			0\ (small)		&   0/small &  0
\\ \ensuremath{m_{CP}^2} (\gev)^2&0& 0   & 0 & 0 & 42^2
\\ \hline	
 \sin^2 \theta_{sh} & 0.1 & 10^{-3} & .22		& 0.06&  .10
\\ \sin^2 \theta_{ah}&0 & 0& 0	&	0		& .01\ (.01)
\\ m_{\tilde{h}}(\gev) &109 & 94 &  94.9	& 111& 103
\\ m_{\tilde{s}} (\gev)&73.8 & 503 & 76.2		&39.3&67.0\ (66.6)
\\ m_{\tilde{a}} (\gev)& 32.6 & 38.7 (11.1) & 28.2\ (8.37)		& 16.2\ (7.13) &18.4\ (9.87)
\\ \hline
 B_{\fh \to 2\fa}& 0.86 & 0.92 (0.96) & .92\ (.93)		& .35\ (.36) &.09\ (.12) 
\\  B_{\fh \to 2\fs}& 0 & 0  &  0       		& .50\ (.49) & [\fa\fs]\, .70\  (.69)
\\ B_{\fs \to 2\fa}&.99 &  0 & .99		& .99\ (.99)  & .91\ (.96)
\\ a\to X & bb & gg & 	gg\ (\tau \tau) &gg, bb\ (\tau\tau) & bb\ (\tau \tau)
\\ \hline \rm tuning &3\% &  60\% (3\%)& 100\%\ (10\%) 		& 4\% (2\%)  & 100\%\ (18\%)
\\ \rm pheno &\rm i & \rm iii & \rm ii, iii & \rm v & \rm iv
\\ \hline 
\end{array} \end{math}\nonumber
\caption{Benchmark Point Summary. Terms in parentheses are variations on the spectrum in order to achieve $\fa \to \tau \tau$ decays. When $A_h$ is labeled 0/small, the 0 indicates the presence of nonzero $M_Q^{-1}$ allowing $\fa \to gg$ decays, while the ``small'' is to indicate a small level allowing $\fa \to bb$ or $\fa \to \tau \tau$ decays, without modifying the spectrum. [$\tilde a \tilde s$] indicates that the decay is $\fh \to \fa\fs$ rather than $\fh \to 2\fs$. \label{table:benchmark}}}

\subsection{Possible exclusions in existing data}
To the extent that many of these signals have not been explicitly analyzed, one can argue that only the model independent bound from OPAL truly limits them. However, once one specifies a given decay mode, it is almost certain that the bounds will improve from the model independent limit. 

Going further, extrapolating from existing limits might give us an estimate on the
potential exclusions of an actual analysis.  For instance, in searches for $h\to 4b$ in
events where the Z decays hadronically, DELPHI and OPAL force the whole event into 4 jets,
which is reasonably efficient even though the total process can have more jets in it (up
to 6).  So since this analysis is akin to the SM 4 jet analysis done for $h\to 2b,  Z \to
{\rm hadronic}$, the flavor-independent analyses may be used to estimate the potential
limits on $h\to 4g$.  If one assumes that the efficiencies to
reconstruct the $4g$ and $4b$ state are the same, and that the background for $4g$ is the same as for $2g$, one can estimate the
scale of exclusion that may be possible in existing data.  Doing this, it appears that $h\to 4j$ below 86 GeV has strong exclusion, that interesting constraints could possibly be set for $86\ \gev < m_{\tilde h} < 100\ \gev$, and for $m_{\tilde h}>100 \gev$ it appears
unlikely that interesting constraints could be set. All of this is an extremely rough estimate,
but suggests additional analyses in the hadronic channels should be done.

Decays of the Higgs to six and eight parton final states are more difficult to estimate, for instance it is not known how sensitive the current $2b$ and $4b$ searches are to actual $6b$ or $8b$ decays.  Analyses that attempt to utilize the multi-jet nature of the decays might be useful, but might encounter issues like jet-finding algorithms constructing fake jets in the signal or background (OPAL and DELPHI use the DURHAM algorithm which is known to have such issues).  It is also not clear how much b-tagging can help in the cases of multi-b decays.

\subsection{Future experiments}
It is not clear to what extent these Higgses may evade detection at the LHC. The six and eight parton decays almost certainly will be challenging. Similarly, the decay of the Higgs into four gluon jets is a tremendous challenge at any hadron collider. The decay $h\to 2a \to 2g 2 \gamma$ should be more promising. Especially as loops involving the Higgsinos can amplify this beyond the expected $7 \times 10^{-3}$ branching ratio, analyses of such scenarios, particularly at the Tevatron, would be highly motivated.  The associated $h\to 4\gamma$ decay has the added signal of seeing both the two $a$'s and the Higgs that decayed into them.  Unfortunately, the expected $10^{-5}$ branching ratio is probably too small to be seen, so this channel requires either an enhanced rate either through increased branching ratio or Higgs production via colored sparticle decay chains.

\section{Summary and Conclusions}
\label{sec:conclusions}
One of the strongest predictions of the MSSM is the character of the Higgs boson, both in
its mass and its couplings. However, the simplest extension of the MSSM, adding a singlet,
can dramatically alter the Higgs boson phenomenology, in particular by introducing a
remarkable set of decay scenarios.  We have studied, in a relatively general sense, the
effects of singlets on the decays of the Higgs boson in supersymmetric theories. There are
a number of general points to be gleaned from this analysis.

First of all, the MSSM expectations of the Higgs boson decays can easily be modified. The
presence of the singlet opens up a number of cascade decay possibilities, which are much
harder to constrain.  Secondly, the NMSSM (where the singlet acquires an expectation value
to generate the $\mu$-term) is too narrow a framework to realize many of the interesting decay scenarios. A number of them can only dominate the Higgs decay once one includes the possible effects of D-term vevs on the physics of singlets. Some cascades can occur but are remarkably
tuned within the NMSSM, but are not tuned in a broader framework. This makes it essential
that we continue to include these operators in future analyses.

Finally, while these models admit much lighter Higgses, and hence much lighter stops, than
the MSSM, the sensitivity of $m_Z$ is typically inadequate as a measure of tuning. Often
the most severe tuning comes from finding an appropriate spectrum such
that cascades can occur, while still maintaining a large coupling to the Higgs boson. All
analyses of cascade decays must be careful to consider these additional tunings rather than simply
focus on the tuning of $m_Z$.  Also, through mixing, heavier Higgses can be allowed for
lighter top squarks and hence with less tuning of the $m_Z$.

Such issues notwithstanding, it is quite straightforward to achieve
models which are not tuned, and have a Higgs below the MSSM LEP limit.
Cascade decays $\fh\to 2\fa \to 4 b$ allow lighter, more natural
Higgses, but at the expense of typically saturating the LEP bounds, a
tuning of sorts on its own. Plus, these allowed regions will likely be impacted by the finalized LEP limits.  Other cascade decays, in particular $\fh \to 2\fa \to 4 g$ and $\fh \to \fa\fs \to 3\fa \to 6 b$ can occur with
very little tuning of the $\fa$ mass, and light Higgses ($m_\fh>82\
\gev$). Decays with $\tau$ final states ($\fh\to 2 \fa \to 4 \tau$,
$\fh \to \fa\fs \to 3\fa \to 6 \tau$, $\fh \to 2\fs \to 4\fa \to 8
\tau$) seem to be quite tuned, since there is no apparent symmetry to explain the
very light $\fa$ (except in case of PQ axion, see footnote \ref{foot:pqaxion}). On the other hand, two stage cascades $\fh\to
2\fs \to 4\fa \to 8b, 8g$ can occur with moderate tuning.  In our realizations, two stage cascades (those with six- and eight parton final states) require D-term breaking while gluon final states arise with new, heavy fields, although there are possibly additional means to produce this phenomenology.  

The collider tests of these scenarios are largely unexplored.  Firstly, since the model independent decay analysis of OPAL \cite{Abbiendi:2002qp} is the dominant constraint for most of our scenarios, it would be extremely useful for a LEP-wide analysis to be carried out.  LEP may also have been sensitive to some of the new non-standard decays we have introduced, which motivates further investigation by the LEP Higgs search collaborations on these specific decays.  Therefore, we strongly suggest that the following analyses be performed:
\begin{itemize}
\item{$h \to$ anything (model independent). Such an exclusion performed by the OPAL analysis is an important consideration for {\em all} models of non-standard Higgs decays, and thus should be performed by as many LEP collaborations as possible.}
\item{$h \to$ hadronic (i.e., 4 jet, or multi-parton final states). Given that photonic, leptonic, and two jet decays of the Higgs are already highly constrained, it is worthwhile to close a significant means by which a Higgs might hide, namely in a generic hadronic decay. However, such an analysis may be quite difficult to separate from background, even in events where large amounts of $b$-flavor is required.  The simplest analysis should be an extension of flavor-independent Higgs search to flavor-independent cascades.}
\item{$h \to 2a \to 4 \tau$ above $m_{\tilde h}=85\ \gev$. The theoretical considerations which appear to have truncated the analysis at this mass range do not apply in general models, and likely significant constraints can still be placed above this value.}
\end{itemize}

At the LHC and the Tevatron, the most promising decay channel is the $\fh\to 2\fa \to 2g\,
2\gamma$ that is associated with 4 gluon decay. At the Tevatron, such a search may be
possible, because the jet threshold is sufficiently low. At the LHC, in this topology, the
\fa\ photon decay could be seen, although a more careful analysis is needed
\cite{Dobrescu:2000jt}.  Since the two gluon jets would be difficult to measure/choose,
this topology does not allow detection of the Higgs.  The 4 photon decay is probably at
too small of a rate to detect, although this may be increased either by increasing the
branching ratio or if Higgses appear in cascade decays of colored superpartners (and thus
boosting the overall production rate).  Many of the other channels seem even more
difficult to search at the LHC, but on a positive note, at the ILC, one should be able to
detect the Higgs through the Z recoil method (see for example \cite{Janot:1991af} and for a linear collider analysis of a SUSY model with nonstandard Higgs decays see \cite{Han:2004yd}).   

It should be noted that while many of these scenarios are very difficult in terms of Higgs detection, they are not nightmare scenarios. Quite the contrary, the existence of these decays allows a light Higgs and would naturally be associated with a wealth of light superpartners. This suggests that the proper lesson of LEP is not a need for models with a heavy Higgs, but rather that further thinking about the possibilities of a stealthy Higgs is in order.


\acknowledgments
The authors thank Martin Boonekamp, Radovan Dermisek, Rouven Essig, Ian Hinchliffe, David E. Kaplan, Tom Junk, Amit Lath, Konstantin Matchev, Maxim Perelstein, Andre Sopczak, Matt Strassler, Scott Thomas, Jay Wacker and Scott Willenbrock for useful conversations and correspondence.  SC would like to thank Christopher Tully for pointing out the presentation at SUSY 05 \cite{Sopczak:2005su}. The authors would especially like to thank Mark Oreglia for extensive discussions regarding the details of LEP Higgs analyses.   PF would like to thank the CCPP/NYU for kind hospitality while part of this work was completed.  PF and NW thank Technion where this work was initiated.  The authors would also like to thank the Aspen Center for Physics where a portion of this work was completed.  The work of S. Chang and N. Weiner was supported by NSF CAREER grant PHY-0449818. P.J. Fox was supported in part by the Director, Office of Science, Office of High Energy and Nuclear Physics, Division of High Energy Physics, of the US Department of Energy under contract DE-AC02-05CH11231.
\bibliographystyle{JHEP}
\bibliography{decay}

\appendix{
\section{Trilinear couplings}
\label{sec:trilinears}

\subsection*{\ensuremath{\mathbf{m_D}} -- Supersoft Operator}
The terms arising from the supersoft operator are
\bea
\st^3 && \left[-\frac{ c_{2\alpha}
 s^2_{\theta}}{4}(g_Y m_D c_\theta )\right] \\
\st^2 \hht && \left[\frac{c_{ 2\alpha} s_\theta }{4}\left(g_Y m_D
(2 c^2_{\theta} - s^2_{\theta})\right)\right] \\
\st\hht^2 && \left[-\frac{ c_{ 2\alpha} c_\theta }{4}\left(g_Y m_D
( c^2_{\theta}-2 s^2_{\theta})\right)\right] \\
\hht^3 && \left[-\frac{c_{ 2\alpha} c^2_{\theta}}{4}\left(g_Y m_D
 s_\theta \right)\right] \\
\hht\at^2 && \left[-\frac{c_{2\beta}s_{\theta}}{4}(g_Y m_D s_{\phi}^2)\right] \\
\st\at^2 && \left[-\frac{c_{2\beta}c_{\theta}}{4}(g_Y m_D s_{\phi}^2)\right]
\eea
while those arising from the D-terms are

\subsection*{D-terms}
\bea
\st^3  && \left[\frac{ c_{2\alpha} s_{\alpha+\beta}}{8}\left(\sqrt{2}\Delta v  s^3_\theta \right)\right] \\
\st^2\hht && \left[-\frac{c_{ 2\alpha} s_{\alpha+\beta}}{8}\left( 3\sqrt{2}\Delta v
  s^2_\theta c_\theta \right)\right] \\
\st\hht^2 && \left[\frac{c_{ 2\alpha} s_{\alpha+\beta}}{8}\left(3\sqrt{2}\Delta v s_\theta c^2_\theta 
  \right)\right] \\
\hht^3 && \left[-\frac{c_{ 2\alpha}
s_{\alpha+\beta}}{8}\left(\sqrt{2}\Delta v c^3_\theta 
\right)\right]\\
\hht \at^2 && \left[-\frac{c_{2\beta}s_{\alpha+\beta}}{8}
\left(\sqrt{2}\Delta v c_\theta s_\phi^2\right)\right]\\
\st\at^2 && \left[\frac{c_{2\beta}s_{\alpha+\beta}}{8}
\left(\sqrt{2}\Delta v s_\theta s_\phi^2\right)\right] 
\eea
where at tree-level $\Delta=g_Y^2+g^2$.  Where the notation is as in
section \ref{sec:singlets}; $\theta$ is the mixing angle between $s$
and $h$ and $\phi$ is the mixing angle between $a$ and $A^0$, from the
\ensuremath{A_h} operator.

\subsection*{Top corrections}
The loops from the top induce trilinear terms with an overall coefficient $C=\frac{3y_t^4}{16 \pi^2} \log(m_{\tilde t_1} m_{\tilde t_2}/m_t^2)$
\bea
\st^3   && \left[\sqrt{2} v s_\theta^3 c^3_\alpha s_\beta C \right]\\
\st^2 \hht   && \left[- 3 \sqrt{2}v c_\theta s_\theta^2 c_\alpha^3
s_\beta C \right]\\
\st\hht^2   && \left[3 \sqrt{2} v c_\theta^2 s_\theta c_\alpha^3
s_\beta C \right]\\
\hht^3   && \left[- \sqrt{2} v c_\theta^3 c_\alpha^3 s_\beta C \right] \\
\hht\at^2   && \left[-\sqrt{2} v c_\theta c_\alpha s_\beta c_\beta^2 s_\phi^2 C \right] \\
\st\at^2   && \left[\sqrt{2} v s_\theta c_\alpha s_\beta c_\beta^2 s_\phi^2 C \right] 
\eea

\subsection*{\ensuremath{\mathbf{\lambda_h SH_u H_d}}}
The $\lambda_h$ and $\mu$ term give trilinears
\bea
\st^3  && \left[-\frac{\lambda_h \tilde{\mu} s_\theta^2 c_\theta}{\sqrt{2}} - \frac{\lambda_h^2 v}{\sqrt{2}} (s_\theta c_\theta^2 s_{\alpha-\beta}+\frac{1}{2} s_\theta^3 s_{2\alpha} c_{\alpha+\beta})\right] \\
\st^2 \hht  && \left[-
\frac{\lambda_h \tilde{\mu}}{\sqrt{2}}s_\theta (s_\theta^2-2 c_\theta^2) - \frac{\lambda_h^2 v}{\sqrt{2}} c_\theta s_{\alpha-\beta}(2s_\theta^2-c_\theta^2) + \frac{3\lambda_h^2 v}{2\sqrt{2}}s_\theta^2 c_\theta s_{2\alpha} c_{\alpha+\beta}\right] \\
\st \hht^2 && \left[-\frac{\lambda_h \tilde{\mu}}{\sqrt{2}}c_\theta (c_\theta^2-2s_\theta^2) - \frac{\lambda_h^2 v}{\sqrt{2}}s_\theta s_{\alpha-\beta}(s_\theta^2-2 c_\theta^2) - \frac{3\lambda_h^2 v}{2\sqrt{2}}c_\theta^2 s_\theta s_{2\alpha}c_{\alpha+\beta}\right] \\
\hht^3 && \left[- \frac{\lambda_h\tilde{\mu}}{\sqrt{2}} c_\theta^2 s_\theta + \frac{\lambda_h^2 v}{\sqrt{2}} (s_\theta^2 c_\theta s_{\alpha-\beta}+\frac{1}{2}c_\theta^3 s_{2\alpha}c_{\alpha+\beta})
\right] \\
\hht \at^2 && \left[-\frac{\lambda_h\tilde{\mu}}{\sqrt{2}}s_\theta s_\phi^2 +\frac{\lambda_h^2 v}{\sqrt{2}} (c_\theta c_\phi^2 s_{\alpha-\beta}+c_\theta s_\phi^2 (s_\alpha c_\beta^3-c_\alpha s_\beta^3))
\right] \\
\st \at^2 && \left[-\frac{\lambda_h^2 \tilde{\mu}}{\sqrt{2}} c_\theta  s_\phi^2- \frac{\lambda_h^2 v}{\sqrt{2}} (s_\theta c_\phi^2 s_{\alpha-\beta}+s_\theta s_\phi^2(s_\alpha c_\beta^3-c_\alpha s_\beta^3))
\right]
\eea

\subsection*{\ensuremath{\mathbf{A_s}}}
In this case the induced trilinears are
\bea 
\st^3 && \left[- \frac{A_s}{\sqrt{2}} c^3_\theta 
\right] \\
\hht\st^2 && \left[-\frac{3}{\sqrt{2}} A_s  c_\theta^2 s_\theta
 \right] \\
\hht^2\st && \left[-\frac{3}{\sqrt{2}} A_s c_\theta s^2_\theta  
\right] \\
\hht^3  && \left[- \frac{A_s}{\sqrt{2}}  s^3_\theta
\right]\\
\hht \at^2  && \left[\frac{3}{\sqrt{2}} A_s s_\theta c_\phi^2\right] \\
\st \at^2  && \left[\frac{3}{\sqrt{2}}A_s c_\theta c_\phi^2 \right] 
\eea

\subsection*{\ensuremath{\mathbf{A_h}}}
This soft term gives the following trilinears 
\bea
\st^3 && \left[\frac{A_h}{2\sqrt{2}} s^2_\theta c_\theta s_{2\alpha} 
\right] \\
\st^2\hht && \left[\frac{A_h}{2\sqrt{2}} s_\theta s_{2\alpha}
(s_\theta^2-2 c_{\theta}^2) 
\right] \\
\hht^2\st && \left[-\frac{A_h}{2\sqrt{2}} c_\theta s_{2\alpha}
(2s_\theta^2-c_\theta^2)
\right]\\
\hht^3 && \left[\frac{A_h}{2\sqrt{2}}s_\theta c^2_\theta s_{2\alpha}
\right] \\
\hht\at^2 && \left[\frac{A_h}{\sqrt{2}}s_\phi (\frac{1}{2} s_\theta s_\phi s_{2\beta}+c_\theta c_\phi s_{\alpha-\beta})
\right]\\
\st\at^2 && \left[\frac{A_h}{\sqrt{2}} s_\phi(\frac{1}{2} c_\theta s_\phi s_{2\beta}-s_\theta c_\phi s_{\alpha-\beta})
 \right] 
\eea

\subsection*{\ensuremath{\mathbf{m_{CP}^2}}}

Since we are ultimately interested in $\fh\to \fs\fa$ and $\fs \to \fa\fa$ decays, we find those terms in the Lagrangian. Arising from the usual MSSM D-terms, we have (at leading order in $m^2_{CP}$),
\bea
\tilde a \tilde s \tilde h  && \left[-\frac{3 \sqrt{2} m^2_{CP} (m_{\fh}^2
- m_\fs^2) v \Delta c_{ 2 \alpha} s_{\alpha+\beta} s^2_{2 \theta}}{16
(m_\fs^2 - m_a^2)(m_\fh^2 - m_a^2)}
\right ]\\
\tilde a^2 \tilde s 
 && \left[- \frac{3 \sqrt{2} \Delta v m^4_{CP} c_{ 2 \alpha} c^2_\theta
s^3_\theta  s_{\alpha+\beta}(m_\fh^2 - m_\fs^2)^2  }{8
(m_\fs^2-m_a^2)^2(m_\fh^2-m_a^2)^2} \right ]
\eea
While from \ensuremath{m_D} we have
\bea
\tilde a \tilde s \tilde h  && \left[ \frac{m^2_{CP} g_Y m_D c_{ 2 \alpha }(-2 m_a^2 + m_\fh^2+m_\fs^2 + 3 (m_\fh^2 - m_\fs^2) c_{ 2 \theta} ) s_{2\theta}}{8 (m_\fh^2 - m_a^2)(m_\fs^2 - m_a^2)} \right ] \\
\tilde a^2 \tilde s  && \left[-\frac{m^4_{CP} g_Y m_D  (m_\fh^2 - m_\fs^2) c_{2 \alpha} s^2_\theta (-4 m_a^2 + 3 m_\fh^2+ m_\fs^2 + 3 (m_\fh^2 - m_\fs^2) c_{2 \theta} )c_\theta}{8 (m_\fh^2 - m_a^2)^2 (m_\fs^2 - m_a^2)^2} \right ] 
\eea
This term is often important in conjunction with \ensuremath{A_s}. The equivalent trilinears with the combination of \ensuremath{A_s} and \ensuremath{m_{CP}^2} are
\bea
 \fa \fs \fh  && \left[ 6 A_s m^2_{CP}  c_\theta  s_\theta \left(\frac{c^2_{ \theta}}{m_\fs^2 - m_a^2}+ \frac{ s^2_{\theta}}{m_\fh^2-m_a^2} \right) \right] \\
\fa^2 \fs  && \left[ 3 A_s m^4_{CP}  c_\theta  \left(\frac{c^2_{ \theta}}{m_\fs^2 - m_a^2}+ \frac{ s^2_{\theta}}{m_\fh^2-m_a^2} \right)^2 \right] 
\eea

}
\end{document}